# A data fusion approach for mobility hub impact assessment and location selection: integrating hub usage data into a large-scale mode choice model

Xiyuan Ren*, Joseph Y. J. Chow

C2SMART University Transportation Center, New York University Tandon School of Engineering, Brooklyn, USA
* Corresponding author: xr2006@nyu.edu

**Abstract**

As cities grapple with traffic congestion and service inequities, mobility hubs offer a scalable solution to align increasing travel demand with sustainability goals. However, evaluating their impacts remains challenging due to the lack of behavioral models that integrate large-scale travel patterns with real-world hub usage. This study presents a data fusion approach that incorporates observed mobility hub usage into a mode choice model estimated with synthetic trip data. We identify trips potentially affected by mobility hubs, introduce a nested choice structure that accounts for mode transfers, and calibrate hub-specific parameters using on-site survey data and ground truth trip counts. A sensitivity analysis demonstrates that the calibration remains robust when the share of hub trips is relatively low and travel demand spans multiple OD pairs. We apply this approach to a case study in Capital District, NY, using data from a survey conducted by the Capital District Transportation Authority (CDTA) and a mode choice model estimated with Replica Inc.'s synthetic data. A bootstrap procedure quantifies uncertainty in hub usage and all downstream impact estimates. The two implemented hubs — near UAlbany Downtown Campus and in Downtown Cohoes — are projected to generate 9.89 (95% CI: [3.20, 29.04]) and 6.98 ([3.86, 12.67]) multimodal trips per day, reduce daily vehicle-miles-traveled (VMT) by 43.29 ([11.69, 142.81]) and 32.12 ([1.04, 60.27]) miles, and increase daily consumer surplus by $3,870 ([2,020, 5,276]) and $1,790 ([1,202, 2,068]), respectively. A regional evaluation of 1,100 candidate locations highlights that optimal hub siting varies by planning objective, with hubs along intercity corridors and urban peripheries yielding the largest behavioral impacts.

## 1. Introduction

Rapid urbanization and growing travel demand have intensified challenges such as traffic congestion, increased carbon emissions, and exacerbated services disparities, significantly impacting urban sustainability and livability (Ecer et al., 2023). As transportation continues to be a major contributor to environmental degradation, innovative solutions that enhance multimodal integration and reduce reliance on single-occupancy vehicles are urgently needed (Hensher et al., 2021). Mobility hubs emerge as critical interventions in this context, offering nodes of connectivity that seamlessly combine public transit, active mobility, and shared services to bridge gaps between disparate modes and foster inclusive urban mobility networks (So et al., 2023). Strategically positioned at key transit locations, mobility hubs encourage transit ridership, reduce car usage, and facilitate convenient transfers in underserved areas (Miramontes et al., 2017), serving as a scalable policy tool to promote sustainable, accessible, and equitable systems that effectively respond to diverse and evolving community needs (Ku et al., 2022).

Considering these potential benefits, selecting locations for mobility hubs, i.e., determining optimal sites from a finite set of candidates based on multiple criteria evaluated simultaneously, becomes an essential and strategic task for policymakers and relevant authorities. Existing studies on hub location selection either employ Multi-Criteria Decision Making (MCDM) methods (Aydin et al., 2022) or formulate optimization models for hub location problems (HLPs) (O'Kelly et al., 2025; Wang et al., 2024). Metrics such as accessibility, environmental impact, and cost-effectiveness have been widely used (So et al., 2023). However, these studies highly depend on the assumed demand, expert judgment, or proxy indicators. An empirical assessment of how mobility hubs impact travelers' mode choices is lacking. Although Xanthopoulos et al. (2024) integrated travel utility from a mode choice model into an optimization framework to determine mobility hub location and capacity in Amsterdam, inadequate consideration of hub-specific parameters, heterogeneities in user preferences, and intercity travel patterns constrain the scalability of their analyses.

Estimating behavioral models that account for both large-scale travel patterns and user preferences of mobility hubs is challenging. On the one hand, citywide or regional travel data typically do not include multimodal trips involving the use of mobility hubs since most hubs are still in the conceptual stage, and few have been implemented (Arnold et al., 2023; Rongen et al., 2022). The lack of real-world usage data makes it difficult to obtain user feedback and accurately capture user preferences regarding mobility hubs. On the other hand, hub usage data from on-site investigations are often limited in both spatial and temporal scope (Horjus et al., 2022). Mobility hubs involve diverse mode combinations as well as numerous potential trips (Arseneault, 2022), which cannot be fully represented by surveys with small sample sizes.

This study proposes a novel data fusion approach to integrate mobility hub usage data into a mode choice model estimated with large-scale information and communication technology (ICT) data. The mode choice model quantifies mode shares for a wide range of trip origins to destinations (OD) pairs, segmented by age, income level, and employment status. The hub usage data provide direct insights into how travelers use mobility hubs, enabling the calibration of hub-related parameters not included in the initially estimated model. However, such data alone would not provide an adequate perspective of modal patterns in the greater region that include users who choose not to use the hubs.

The data fusion approach involves three key steps. First, we identify trips potentially impacted by mobility hubs and generate a revised choice set that includes multimodal trip options. Second, we extend a choice model for unimodal travel modes by introducing a nested choice structure that explicitly accounts for mode transfers enabled by mobility hubs. Third, we calibrate hub-specific parameters within the pre-estimated choice model using survey and ground-truth data. With this hub-usage-enhanced modeling framework, it becomes possible to quantify the behavioral impacts of mobility hubs, including changes in transit ridership (Wu & Liao, 2020), vehicle miles traveled (VMT) (Shin, 2020), carbon emission (Hosseini et al., 2024), and consumer surplus in the post-pilot scenario (McHardy et al., 2023).

In a case study, we apply this method to assess the impacts of two mobility hubs in the State University of New York (SUNY), Albany (UAlbany), and Cohoes, NY. The Capital District Transportation Authority (CDTA) initiated the Capital Region Mobility Demonstration Project from April 2022 to June 2024 to test out two mobility hubs (CDTA, 2024). An on-site survey was designed to collect hub usage data, resulting in 40 complete responses gathered from October to December 2023. CDTA also provided backend data of DRIVE (a car-share service available at the mobility hubs) and transit ridership to and from those hub stations. Clearly, this low sample would not be adequate for building a model on its own. Data fusion allows this small sample to be used in a supplemental manner that builds on the strength of an existing model.

For the pre-trained choice model, we utilized a New York State mode choice model estimated using synthetic trip data provided by Replica Inc. The model, publicly available at https://zenodo.org/record/8113817, includes agent-specific parameters accounting for statewide taste heterogeneity and mode shares for census-block-group-level OD pairs (see Ren et al. (2025) for the methodology behind the model). The impact of the mobility hub demonstration is measured by comparing post-pilot demand and usage of the two hubs against a pre-deployment baseline. We extend the assessment to all potential hub locations within the Albany-Schenectady-Troy Metropolitan Statistical Area (MSA). These locations are ranked based on potential demand, increased transit ridership, reduced vehicle miles traveled (VMT), and increased trip consumer surplus (CS) (McConnell, 1995).

The remainder of the paper is organized as follows: Section 2 outlines mobility hub deployment strategies, choice modeling techniques, and our contributions. Section 3 presents our data fusion approach with an illustrative example demonstrating how it works. Section 4 reports the results of an empirical study in New York's Capital District, where a mobility hub demonstration project took place from 2022 to 2024. Section 5 offers conclusions and future research directions.

## 2. Literature review

### 2.1. Metrics for hub assessment and hub location problems (HLPs)

Existing studies identify three key advantages of mobility hubs: the revitalization of public transportation, environmental benefits, and support for multimodal travel. Environmentally, mobility hubs mitigate emissions by shifting trips from private vehicles to shared and public modes, particularly when intermodal connections are optimized for efficiency—a process sometimes referred to as "boosting metabolism" (Arias-Molinares et al., 2023; Sihvonen & Weck, 2023). In addition, hubs enable flexible multimodal journeys, combining options such as park-and-ride, bus-bike transfers, or shared micro-mobility to improve travel efficiency and reduce overall costs (McHardy et al., 2023; Weustenenk & Mingardo, 2023).

Given these potential benefits, a growing body of literature has focused on developing corresponding metrics for hub assessment. Traditional evaluations often rely on aggregate indicators, such as total travel time, generalized cost reduction, or accessibility gain, to capture how hubs improve overall system performance (Sun et al., 2019). More recent studies extend these measures through network-level metrics, including hub centrality, connectivity, and utilization, which evaluate the structural role of hubs in facilitating multimodal transfers and enhancing regional reach (Rongen et al., 2022). Sustainability-oriented indicators, such as greenhouse gas reduction, increased transit ridership, and decreased private car usage, have gained growing attention in both research and practice (Hansel, 2025; Tran & Draeger, 2021). Another stream of research broadens the scope by integrating resilience metrics, such as the ability of hub networks to maintain connectivity under disruption (Wen et al., 2025), and equity-oriented measures, such as assessing whether disadvantaged groups benefit from improved access (Martinez et al., 2024).

Hub locations problems (HLP) framework originates from O'Kelly (1986) and subsequent formulations, such as p-hub median, p-hub center, and hub covering problems (Wang et al., 2024). These models seek to determine optimal hub locations and allocation patterns to minimize total transportation cost while exploiting economies of scale through hub-and-spoke consolidation (Wandelt et al., 2022). Variants include the single-allocation and multiple-allocation structures, capacitated and uncapacitated models, and formulations considering facility costs, congestion, and uncertainties. A comprehensive meta-review by O'Kelly et al. (2025) synthesized more than three decades of research, identifying emerging challenges related to data quality, robustness, temporal aspects, integration of emerging modes, and interdisciplinary approaches. These challenges signal the diversification of the HLP field from purely operational optimization to multidimensional applications. Aligning with the calls for insight-based research rather than purely numerical evaluation of hub design outcomes (O'Kelly et al., 2025), several studies have adopted multi-criteria decision-making (MCDM) methods to incorporate a broader set of metrics. For instance, Aydin et al. (2022) introduced an MCDM framework that integrates the Analytic Hierarchy Process (AHP) and Weighted Aggregated Sum Product Assessment (WASPAS) methods to evaluate mobility hub locations on the Anatolian side of Istanbul. So et al. (2023) proposed a two-step methodology—combining feasibility analysis with AHP—to identify optimal hub locations in Seoul.

However, a key limitation persists: most existing works simplify demand-side behavioral dynamics, particularly how traveler preferences and mode choices respond to hub availability. Trip demand for mobility hubs is often estimated using proxy indicators such as population density (Miramontes et al., 2017) or derived from equilibrium frameworks that consider only travel cost of hub-related modes (Frank et al., 2021). Although Xanthopoulos et al. (2024) present one of the few studies that based on travelers' trip utilities, hub-specific parameters in their studies are assumed to be homogeneous and are determined based on expert judgment. The oversimplification of behavioral preferences can distort the estimation of key evaluation metrics—such as travel demand, trip utility, and environmental benefit—and underestimate the trade-offs among them during hub site selection.

### 2.2. Discrete choice models (DCMs) for hub impact assessment

DCMs are widely applied in transportation research to estimate travel demand by assuming individuals make choices by maximizing the overall utility they can expect to gain (see Bowman & Ben-Akiva, 2001). These models enable researchers to examine how various factors, including cost, travel time, and personal preferences, influence the likelihood of selecting specific modes of transportation (Greene & Hensher, 2003; Hensher & Ho, 2016). According to the mixed logit (MXL) framework (McFadden & Train, 2000), the utility and probability of individual $n$ choosing alternative $j$ are defined as $U_{nj} = \beta^T X_{nj} + \varepsilon_{nj}$, where $U_{nj}$ is the overall utility of individual $n$ choosing alternative $j$, which consists of a systematic utility $\beta^T X_{nj}$ and a random utility $\varepsilon_{nj}$ usually assumed to be independent and identically distributed (i.i.d.). $X_{nj}$ denotes a set of observed attributes of alternative $j$ for individual $n$. $\beta$ is a vector of taste parameters assumed to vary randomly across individuals to reflect taste heterogeneity.

A number of studies proposed semi-parametric or nonparametric approaches to capture taste heterogeneity in a more flexible manner. For example, Train (2016) proposed a logit-mixed logit (LML) model, in which the mixing distribution of parameters can be easily specified using splines, polynomials, or any other functional forms. Swait (2022) introduced the individual parameter logit (IPL), where individual taste parameters are estimated in a distribution-free manner under the objective of maximizing sample posterior loglikelihood. Ren and Chow (2022) proposed an agent-based mixed logit (AMXL) model using a hybrid machine learning and econometric approach to estimate deterministic, individual-specific parameters. These approaches capture heterogeneous user preferences by estimating agent-level parameters, which provide new opportunities for impact assessment based on mode choice predictions.

Mobility hubs provide a range of new options and services, such as shared bikes, scooters, car rentals, and public transit connections, all of which can shape travelers' decisions and impact trip utilities (Arnold et al., 2023). As such, DCMs are well-suited for assessing the behavioral impacts of mobility hubs, particularly in quantifying travel demand, shifts in mode choices, changes in VMT, and user satisfaction (Saltykova et al., 2022). Despite their methodological relevance, few studies have applied emerging choice modeling techniques to large-scale evaluations of mobility hubs. This gap is largely due to two key challenges: (1) citywide or regional travel datasets typically lack detailed information on multimodal trips, especially those involving hub-based transfers (Rongen et al., 2022), and (2) hub usage data collected through on-site surveys are often limited in spatial and temporal coverage and cannot fully capture the diversity of user behavior (Horjus et al., 2022). Addressing these limitations through integrated data fusion and advanced modeling approaches is crucial for scalable and behaviorally grounded impact assessments of mobility hubs.

### 2.3. Our contributions

This study presents a data fusion framework that integrates large-scale travel data, advanced choice modeling, and real-world mobility hub usage to support behavior-oriented hub impact assessment and location planning. The proposed data fusion framework is operationalized through three methodological steps.

First, we integrate synthetic trip data and on-site hub survey data to identify potential trips influenced by hubs and to construct the corresponding demand and mode choice set. This data fusion approach captures both modeled travel behavior at a regional scale and observed multimodal transfer behavior at the local (hub) scale, enabling a more complete estimation of the total demand affected by hub interventions.

Second, we extend a choice model for traditional travel modes by introducing a nested choice structure that explicitly accounts for mode transfers enabled by mobility hubs. This enhancement allows the model to represent the utility of hub-based mode transfers, such as park-and-ride, bike-to-transit, or bus-to-bus trips—thus bridging conventional mode choice modeling with the emerging paradigm of integrated mobility.

Third, we calibrate hub-related parameters using ground-truth trip counts and observed mode shares, allowing the model to simultaneously capture heterogeneity in travelers' preferences and hub-specific utilities. This calibration ensures that the model not only reproduces realistic behavioral responses to travel time and cost but also reflects systematic utilities related to hub usage and mode transfer.

Collectively, the proposed data fusion framework makes three key contributions. (1) It enables a comprehensive assessment of hub impacts on travel demand, mode share, and trip utility by leveraging limited survey data complemented by large-scale synthetic trips; (2) It provides a scalable tool for comparing alternative hub location strategies under multiple evaluation metrics such as potential demand, user welfare, or environmental benefit; and (3) it establishes a robust behavioral foundation for subsequent optimization models, where calibrated taste parameters can be incorporated to simulate equilibrium responses in hub location problems.

## 3. Proposed methodology

This section introduces the proposed data fusion framework designed to integrate large-scale travel data, observed hub usage, and a pre-trained choice model for behavior-oriented mobility hub assessment. An illustrative example is provided to demonstrate how the framework operates and to examine the sensitivity of the results to the availability of hub usage data. In addition, this section defines the evaluation metrics used for assessing hub impacts and hub location selection.

### 3.1. Data fusion framework

Fig. 1 shows the proposed fusion framework, which consists of three major steps: (1) potential trip identification and mode choice set generation, (2) extended mode choice model, and (3) hub-related parameter calibration. Notations used in this section are listed in Table 1.

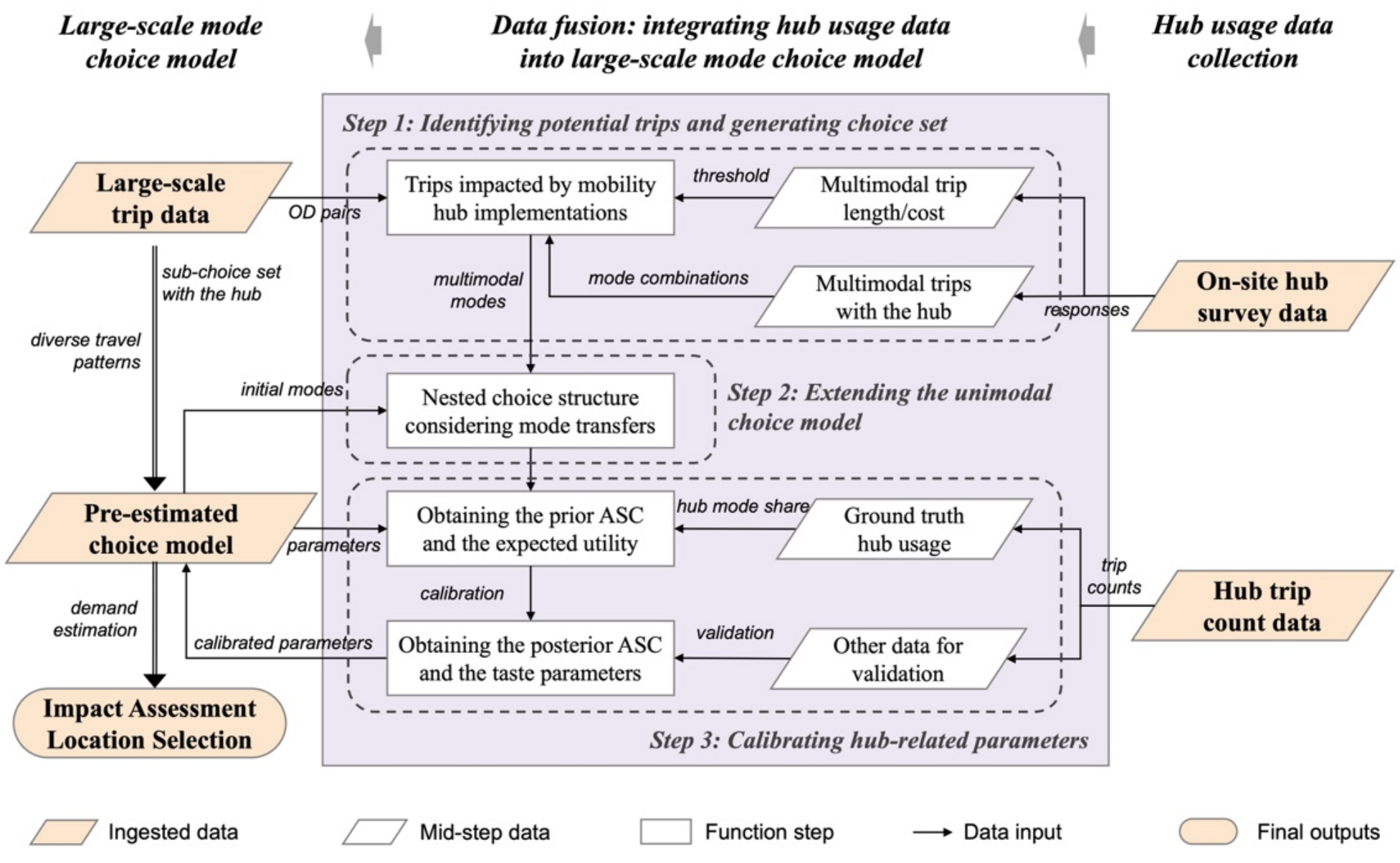

**Fig. 1.** Proposed data fusion framework.

**Table 1**

Notations used in the proposed model

| Notation | Description |
| --- | --- |
| *Sets* | |
| $I$ | Set of population segments, indexed by i |

| $T$ | Set of potentially impacted trip origin–destination (OD) pairs, indexed by t |
|---|---|
| $J$ | Set of traditional (unimodal) mode alternatives, indexed by j |
| $J^+$ | Upper-branch choice set, $J^+ = J \cup \{hub\}$ |
| $H$ | Set of mobility hubs, indexed by h |
| $J_{h_t}^{hub}$ | Set of feasible multimodal alternatives at the hub assigned to OD pair t |
| $(j_1, j_2)$ | A mode pair: first-leg (origin→hub) and second-leg (hub→destination) |
| *Utilities and taste parameters* | |
| $U_{i,j,t}$ | Total utility of mode j for segment i along OD pair t |
| $V_{i,j,t}$ | Systematic utility of mode j for segment i along OD pair t |
| $\varepsilon_{i,j,t}$ | Independent and identically distributed random error term |
| $X_{i,j,t}$ | Vector of attributes of mode j along OD pair t for segment i |
| $\beta_{i,t}$ | Vector of pre-trained taste parameters of segment i along OD pair t |
| $V_{i,hub,t}$ | Systematic utility of the composite hub alternative |
| $V_{i,(j_1,j_2),t}$ | Systematic utility of mode pair (j1, j2) at the hub |
| $V_{i,j_1,O_t \rightarrow h_t}$ | Systematic utility of the first trip segment (mode j1, origin→hub) |
| $V_{i,j_2,h_t \rightarrow D_t}$ | Systematic utility of the second trip segment (mode j2, hub→destination) |
| $\beta_{hub}$ | Nesting parameter: degree of correlation among multimodal options |
| $asc_{i,h_t}^{hub}$ | Segment- and hub-specific hub constant (attractiveness of hub h to segment i) |
| $asc_{i,(j_1,j_2)}^{trans}$ | Segment i's (dis)utility of transferring from mode j1 to mode j2 |
| *Choice probabilities* | |
| $P_{i,t}(j \mid J)$ | Probability of choosing traditional mode $j$ before hubs are available |
| $P_{i,t}(hub \mid J^+)$ | Probability of choosing multimodal alternatives at the upper branch |
| $P_{i,t}\left((j_1, j_2) \mid J_{h_t}^{hub}\right)$ | Probability of choosing multimodal alternative $(j_1, j_2)$ at the lower branch |
| $P_{i,t}\left((j_1, j_2) \mid J \cup J_{h_t}^{hub}\right)$ | Probability of choosing multimodal alternative $(j_1, j_2)$ given the entire choice set |
| $P_{i,t}(j_1 \mid (j_1, j_2))$ | Share of multimodal alternative $(j_1, j_2)$ attributed to its first leg's mode $j_1$ |
| $P_{i,t}(j \mid J^+)$ | Probability of choosing traditional mode at the upper branch |
| $P_{i,t}\left(j \mid J \cup J_{h_t}^{hub}\right)$ | Probability of choosing traditional (unimodal) mode after hubs are available |
| *Trip geography and travel demand* | |
| $OD$ | Linear distance between the trip origin and destination |
| $OH_{t,h}$ | Linear distance from the origin of OD pair t to hub h |
| $HD_{t,h}$ | Linear distance from hub h to the destination of OD pair t |
| $h_t$ | Nearest hub assigned to OD pair t |
| $l_{j_1,O_t \rightarrow h_t}$ | Travel distance of the first leg (mode j1, origin→hub) |
| $l_{j_2,h_t \rightarrow D_t}$ | Travel distance of the second leg (mode j2, hub→destination) |
| $l_t$ | Driving network distance of OD pair t |
| $d_{h,i,t}$ | Trip volume by segment i along OD pair t that is potentially impacted by hub h |
| $Num_h$ | Observed number of trips that used hub h |

*3.1.1. Potential trip identification and mode choice set generation*

Let OD denote the linear distance between the trip origin and destination, OH the distance from the origin to the mobility hub, and HD the linear distance from the hub and the destination. As noted in previous studies, multimodal trips involving a mobility hub are typically considered only if they do not significantly increase the total trip length (Arnold et al., 2023; Rongen et al., 2022). Therefore, a trip OD pair is identified as potentially impacted by a mobility hub if it satisfies one of the following conditions: (1) OH + HD < threshold × OD, or (2) OH + HD < OD + 1km. These conditions are illustrated in Fig. 2. Origins, destinations, and trip volumes can be obtained from synthetic trip data, which are typically generated through the integration of census data and travel surveys to capture travel patterns at a regional scale (Hörl & Balac, 2021). The threshold value used for identification can be derived from on-site hub survey data. For each survey response, the ratio (OH+HD) / OD is calculated using the reported trip origin, destination, and hub location. The threshold is then defined as the 90th percentile (or another value determined by the researcher) of these ratios across all responses.

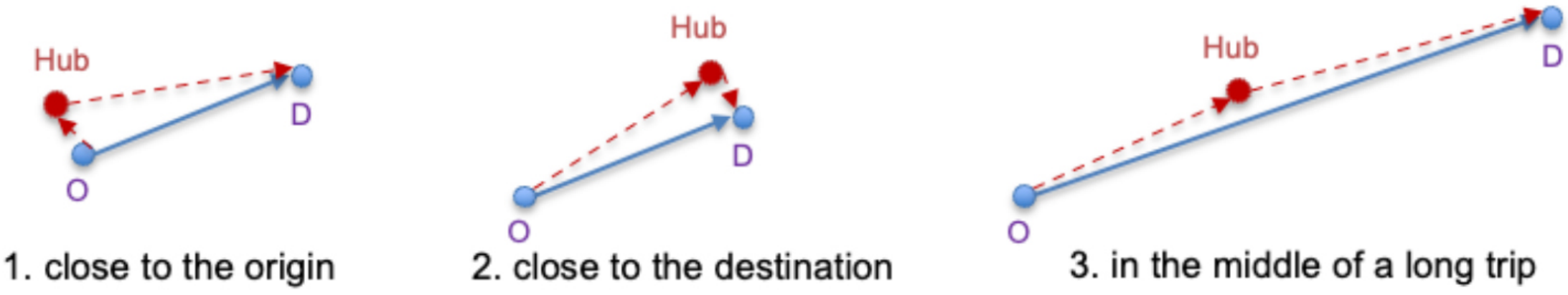

**Fig. 2.** Illustration of trips potentially impacted by a mobility hub.

The entry and exit modes reported in survey responses reflect available mode transfers at the hub. For example, if a traveler took a bus to enter the hub and rode a bike to exit it, the combination "Bus–Bike" should be considered as a multimodal trip option. Accordingly, the travel time and cost for this trip option are obtained by summing the travel time and cost of the two trip segments, along with the additional transfer time and cost incurred at the hub. Based on all available mode transfer combinations identified from the survey, we construct a choice set that includes these hub-enabled trip options as a complement to traditional alternatives without hub implementation. Notably, this extended choice set may also include "Bus–Bus" trips, representing cases where travelers use the hub solely for bus-to-bus transfers.

We restrict our analysis to trips involving a single hub transfer, as multiple-transfer trips would considerably increase model complexity and are rare in our case study (see Section 4.1). Accordingly, each potentially impacted OD pair $t \in T$ is assigned to its nearest hub $h_t$ from the set of candidate hubs $H$, as defined in Eq. (1).

$$h_t = \underset{h \in H}{\arg\min}(OH_{t,h} + HD_{t,h}), \qquad \forall t \in T \tag{1}$$

*3.1.2. Extended mode choice model*

We extend the single-mode choice model by introducing a nested structure, where the upper-level choice captures the decision between traditional modes (e.g., driving, public transit, biking, walking) and hub-enabled trip options, while the lower-level choice represents the selection among specific transfer combinations at the mobility hub.

In the pre-estimated mode choice model, trips sharing the same origins and destinations made by the same population segments have a unique set of taste parameters. Let $i$ denote the index of population segments and $t$ denote the index of OD pairs. The utility and choice probability (i.e., mode share) of alternative $j$ are defined in Eqs. (2) – (3).

$$U_{i,j,t} = V_{i,j,t} + \varepsilon_{i,j,t} = \beta_{i,t} X_{i,j,t} + \varepsilon_{i,j,t}, \qquad \forall i \in I, j \in J, t \in T \tag{2}$$

$$P_{i,t}(j|J) = \frac{A_{i,j,t} e^{V_{i,j,t}}}{\sum_{q \in J} e^{A_{i,j,t} V_{i,q,t}}}, \qquad \forall i \in I, j \in J, t \in T \tag{3}$$

where $I$ denotes the set of population segments; $J$ denotes the set of traditional mode alternatives; $T$ denotes the set of impacted OD pairs. $U_{i,j,t}$ is the total utility of mode $j$ in trips made by segment $i$ along OD pair $t$; $V_{i,j,t}$ is the corresponding systematic utility; $X_{i,j,t}$ is a vector of mode $j's$ attributes along OD pair $t$ for segment $i$; $\beta_{i,t}$ is a vector of taste parameters specific to segment $i$ along OD pair $t$; $\varepsilon_{i,j,t}$ is an i.i.d. distributed random error term. $P_{i,t}(j|J)$ represents the probability that segment $i$ chooses mode $j$ from choice set $J$ for trips along OD pair $t$, and $A_{i,j,t}$ is a binary availability indicator equal to 1 if mode $j$ is available to segment $i$ along OD pair $t$, and 0 otherwise.

In the upper branch, we introduce a composite hub alternative (nest) that complements the existing unimodal alternatives. Accordingly, we define $J^+ = J \cup \{hub\}$ as the upper-branch choice set. Since each OD pair $t$ is assigned to its nearest hub $h_t$ (Eq. 1), all hub-related quantities can be indexed by $t$ alone. Following standard nested logit models (Train, 2009), we assume that the random error terms $\varepsilon_{i,j,t}$ are i.i.d. across alternatives in $J^+$. The probability of choosing the hub alternative and its systematic utility are given in Eqs. (4) – (5).

$$P_{i,t}(hub|J^+) = \frac{e^{V_{i,hub,t}}}{\sum_{q \in J^+} e^{V_{i,q,t}}}, \qquad \forall i \in I, t \in T \tag{4}$$

$$V_{i,hub,t} = \beta_{hub} \ln\left( \sum_{(j_1,j_2) \in J^{hub}_{h_t}} \exp\left(\frac{V_{i,(j_1,j_2),t}}{\beta_{hub}}\right) \right) + asc^{hub}_{i,h_t}, \qquad \forall i \in I, t \in T \tag{5}$$

where $J^{hub}_{h_t}$ denotes the set of all mode-pair combinations involving a transfer at hub $h_t$ (the hub assigned to OD pair $t$); $j_1 \in J$ is the mode of the trip segment from the origin to the hub, and $j_2 \in J$ is the mode of the trip segment from the hub to the destination. The systematic utility $V_{i,hub,t}$ is defined as the logsum of the utilities of all feasible mode-pair combinations at the hub, scaled by a nesting parameter $\beta_{hub}$ and shifted by a segment- and hub-specific constant $asc^{hub}_{i,h_t}$.

In the lower-branch, each hub-related trip option consists of two segments: one from the origin to the hub and one from the hub to the destination. The systematic utility of selecting a mode pair $(j_1, j_2)$ at the hub, denoted $V_{i,(j_1,j_2),t}$, is defined in Eq. (6).

$$V_{i,(j_1,j_2),t} = V_{i,j_1,O_t \to h_t} + V_{i,j_2,h_t \to D_t} + asc^{trans}_{i,(j_1,j_2)}, \qquad \forall i \in I, (j_1,j_2) \in J^{hub}_{h_t}, t \in T \tag{6}$$

where $O_t$ and $D_t$ denote the origin and destination of OD pair $t$; $V_{i,j_1,O_t \to h_t}$ is the utility of the first trip segment; $V_{i,j_2,h_t \to D_t}$ is the utility of the second trip segment; and $asc^{trans}_{i,(j_1,j_2)}$ is a constant capturing the (dis)utility of transferring from mode $j_1$ to $j_2$. Accordingly, the probability of choosing a specific mode pair is defined in Eqs. (7) – (8).

$$P_{i,t}\left((j_1,j_2)\middle|J^{hub}_{h_t}\right) = \frac{e^{V_{i,(j_1,j_2),t}}}{\sum_{(q_1,q_2) \in J^{hub}_{h_t}} e^{V_{i,(q_1,q_2),t}}}, \qquad \forall i \in I, (j_1,j_2) \in J^{hub}_{h_t}, t \in T \tag{7}$$

$$P_{i,t}\left((j_1, j_2) \middle| J \cup J_{h_t}^{hub}\right) = P_{i,t}\left((j_1, j_2) \middle| J_{h_t}^{hub}\right) \cdot P_{i,t}(hub|J^+), \quad \forall i \in I, (j_1, j_2) \in J_{h_t}^{hub}, t \in T \tag{8}$$

where $P_{i,t}\left((j_1, j_2) \middle| J_{h_t}^{hub}\right)$ is the lower-branch probability of choosing mode pair $(j_1, j_2)$ conditional on transferring at hub $h_t$, and $P_{i,t}\left(hub \middle| J^+\right)$ denotes the upper-branch probability of choosing to transfer at a hub as defined in Eq. (4). Their product $P_{i,t}\left((j_1, j_2) \middle| J \cup J_{h_t}^{hub}\right)$ represents the joint probability of selecting mode pair $(j_1, j_2)$ via hub $h_t$ among all available unimodal and multimodal trip options.

It is worth noting that the proposed framework treats the hub alternative as parallel and independent from traditional modes in the upper branch. As a consequence, introducing the hub option reduces the choice probabilities of all traditional modes proportionally, exhibiting the independence of irrelevant alternatives (IIA) property that holds across nests in the nested logit framework (Train, 2009). This reflects a fundamental limitation of nested logit: it can capture correlation among alternatives along only one dimension, which in our case is the multimodal-versus-unimodal distinction. Although this dimension is arguably the most relevant for assessing the impact of mobility hubs, other potentially meaningful dimensions of correlation, such as shared modes versus private modes, or motorized versus non-motorized modes, are not captured. Accommodating multiple correlation dimensions would require extending the formulation to more flexible structures such as the inverse product differentiation logit (IPDL) model (Fosgerau et al., 2024). We treat this as a direction for future work in Section 5.2.

*3.1.3. Calibration of hub-related parameters*

The extended choice model introduced in Section 3.1.2 involves three types of hub-related parameters to be calibrated: (1) the nesting parameter $\beta_{hub}$, a value between 0 to 1 that indicates the degree of correlation among multimodal options within the hub nest; (2) the segment- and hub-specific constants $asc_{i,h}^{hub}$, which capture the attractiveness of hub $h$ to segment $i$ due to unobserved factors; and (3) the transfer constants $asc_{i,(j_1,j_2)}^{trans}$, which capture segment $i$'s (dis)utility of transferring from mode $j_1$ to mode $j_2$. In total, the full parameter set comprises $1 + |I| \cdot |H| + |I| \cdot |J|^2$ hub-related parameters, where $|I|$, $|H|$, and $|J|$ denote the numbers of population segments, hubs, and traditional modes, respectively. Calibrating all of these parameters poses a high risk of overfitting in the context of early-stage mobility hub adoption, where observational data for calibration are typically scarce.

To address this limitation, we propose a simplified version of the model based on two assumptions. First, we assume that the transfer (dis)utility within a hub is constant across all mode combinations but may vary across population segments, i.e., $asc_{i,(j_1,j_2)}^{trans} = asc_i^{trans}$ for all $i \in I$ and $(j_1, j_2) \in J_h^{hub}$. Under this assumption, the transfer term is common to all mode pairs within the lower-branch nest and therefore cancels out of the conditional choice probability in Eq. (7). Its effect is instead absorbed into the upper-branch hub constant, which now reflects the combined attractiveness of hub access and transfer (dis)utility. Second, we assume that the upper-branch hub constant varies across population segments but is homogeneous across hubs, i.e., $asc_{i,h}^{hub} = asc_i^{hub}$ for all $i \in I$ and $h \in H$. This is assuming that, at the early adoption stage, mobility hubs in the same region are typically designed to a common service standard (e.g., comparable facilities, signage, and mode availability), so differences in unobserved factors across hubs are likely to be small relative to those across population segments.

Together, these assumptions reduce the number of hub-related parameters from $1 + |I| \cdot |H| + |I| \cdot |J|^2$ to $1 + |I|$, comprising a single nesting parameter $\beta_{hub}$ and $|I|$ segment-specific constants $asc_i^{hub}$. This substantial reduction renders the calibration feasible without a full maximum likelihood estimation (MLE) that would require a substantial sample of individual hub trips, while retaining the core behavioral structure needed to capture mode shifts and welfare changes. We acknowledge that these assumptions are restrictive, and relaxing them would require richer data that are expected to become available as mobility hubs scale up; we return to this point in Section 5.2.

We calibrate the $|I| + 1$ parameters by setting them as decision variables and formulating a least-squares optimization problem that minimize the gap between predicted and observed number of hub trips. The optimization problem is defined in Eqs. (9) – (10).

$$\min_{\beta_{hub}, asc_i^{hub}} \sum_{h \in H} \left( Num_h - \sum_{i \in I} \sum_{t \in T} d_{h,i,t} \cdot P_{i,t}(hub|J^+) \right)^2 \tag{9}$$

subject to:

$$0 < \beta_{hub} \leq 1 \tag{10}$$

where $H$ denotes the set of mobility hubs; $Num_h$ represents the observed number of trips that used mobility hub $h$; $d_{h,i,t}$ denotes the trip volume made by segment $i$ along OD pair $t$ that is potentially impacted by hub $h$; and $d_{h,i,t} \cdot P_{i,t}(hub|J^+)$ denotes the predicted multimodal trip demand. To ensure that the nested structure remains consistent with random utility maximization (RUM), we impose the constraint shown in Eq. (10). We solve the optimization problem using the Generalized Reduced Gradient (GRG) nonlinear algorithm (Lasdon et al., 1974), which is one of the default solver tools. Since the objective function is nonconvex, the GRG nonlinear algorithm has the risk of getting stuck at a local optimum. However, this issue can be addressed by running the algorithm multiple times with randomized initial values.

### 3.2. Illustrative example

This section uses an illustrative example to demonstrate the robustness of the proposed calibration framework under the simplified model. The setup consists of two population segments (lower-income and higher-income travelers), two mobility hubs (Hub 1 and Hub 2), three unimodal travel modes (driving, public transit, and walking), and three hub-enabled transfer options (Car–Bus, Bus–Bus, and Walk–Bus). The systematic utilities associated with these options are manually specified to mimic those from an estimated mode choice model, following standard practice in the literature (Hensher & Ho, 2016; Ren et al., 2024). For example, driving is treated as the reference mode with normalized utility equal to zero, and higher-income travelers are assumed to exhibit stronger preferences for private car use and relatively lower preferences for transit or walking alternatives.

Four experimental scenarios are considered to reflect different numbers of OD pairs and hub adoption levels (Table 2). Scenario 1 represents a single OD pair with low hub demand, where the number of trips for each segment and hub is fixed and only a small fraction of travelers use the hub (approximately 2%). Scenario 2 maintains the same OD structure but increases the proportion of hub-related trips to approximately 20%, representing a higher adoption level for mobility hubs. Scenarios 3 and 4 extend these conditions to multiple OD pairs by varying both unimodal and multimodal utilities and trip volumes across OD pairs to introduce heterogeneity. Specifically, trip volumes and utility parameters in Scenarios 3 and 4 are perturbed by ±{50,100} or ±{0.05, 0.10}, respectively, to generate five OD pairs reflecting spatial variation in accessibility and mode attractiveness. The observed hub trip counts in these multi-OD scenarios are scaled accordingly to ensure the same adoption levels as those in Scenarios 1 and 2.

**Table 2.** Experimental setups in the illustrative example.

| | Scenario 1 | Scenario 2 | Scenario 3 | Scenario 4 |
|---|---|---|---|---|
| **Trip volume ($d_{h,i,t}$)** | | | | |
| Lower-income, Hub 1 | 600 | 600 | 600 $\pm$ {50,100} | 600 $\pm$ {50,100} |
| Lower-income, Hub 2 | 500 | 500 | 500 $\pm$ {50,100} | 500 $\pm$ {50,100} |
| Higher-income, Hub 1 | 400 | 400 | 400 $\pm$ {50,100} | 400 $\pm$ {50,100} |
| Higher-income, Hub 2 | 300 | 300 | 300 $\pm$ {50,100} | 300 $\pm$ {50,100} |
| **Unimodal utility ($V_{i,j,t}$, [Driving, Transit, Walking])** | | | | |
| Lower-income | [0.0, -0.2, -0.4] | [0.0, -0.2, -0.4] | [0.0, -0.2, -0.4] $\pm$ {0.05, 0.10} | [0.0, -0.2, -0.4] $\pm$ {0.05, 0.10} |
| Higher-income | [0.0, -0.5, -0.7] | [0.0, -0.5, -0.7] | [0.0, -0.5, -0.7] $\pm$ {0.05, 0.10} | [0.0, -0.5, -0.7] $\pm$ {0.05, 0.10} |
| **Multimodal utility ($V_{i,(j_1,j_2),t}$, [Car–Bus, Bus–Bus, Walk–Bus])** | | | | |
| Lower-income, Hub 1 | [-0.1, -0.3, -0.5] | [-0.1, -0.3, -0.5] | [-0.1, -0.3, -0.5] $\pm$ {0.05, 0.10} | [-0.1, -0.3, -0.5] $\pm$ {0.05, 0.10} |
| Lower-income, Hub 2 | [-0.2, -0.6, -0.8] | [-0.2, -0.6, -0.8] | [-0.2, -0.6, -0.8] $\pm$ {0.05, 0.10} | [-0.2, -0.6, -0.8] $\pm$ {0.05, 0.10} |
| Higher-income, Hub 1 | [-0.1, -0.3, -0.4] | [-0.1, -0.3, -0.4] | [-0.1, -0.3, -0.4] $\pm$ {0.05, 0.10} | [-0.1, -0.3, -0.4] $\pm$ {0.05, 0.10} |
| Higher-income, Hub 2 | [-0.2, -0.4, -0.5] | [-0.2, -0.4, -0.5] | [-0.2, -0.4, -0.5] $\pm$ {0.05, 0.10} | [-0.2, -0.4, -0.5] $\pm$ {0.05, 0.10} |
| **Observed hub trips ($Num_h$)** | | | | |
| Hub 1 | 21 (2.1%) | 210 (21%) | 105 (2.1%) | 1,050 (21%) |
| Hub 2 | 15 (1.875%) | 150 (18.75%) | 75 (1.875%) | 750 (18.75%) |

To evaluate the robustness of the calibration, we perturb the observed number of hub trips by $\pm$50% in 5% increments to examine the potential impacts of errors or fluctuations in observed hub usage data (e.g., due to survey sampling or temporal variability). For each perturbation level, we execute the least-squares calibration described in Section 3.1.3 ten times with randomized initial values, compute the average values of $\beta_{hub}$ and $asc_i^{hub}$, and check the percentage error in the predicted hub trip counts.

Fig. 3 shows how the calibrated parameters and predicted hub trip counts respond to perturbations in the observed hub usage. As expected, the prediction error in hub trip counts changes almost linearly with the perturbation level, reflecting the design of the calibration objective, which explicitly minimizes the gap between observed and predicted hub usage. The calibrated nesting parameter $\beta_{hub}$ and segment-specific hub constants $asc_i^{hub}$ remain relatively stable in the low-demand settings (Scenarios 1 and 3), where the share of hub-related trips is small. In these cases, the hub alternative competes weakly with unimodal options, so marginal changes in observed hub trips have limited influence on the upper-branch utility. Moreover, the fluctuations in both predicted trips and calibrated parameters are notably smaller in the multiple-OD scenarios (Scenarios 3 and 4) compared to single-OD ones. This occurs because hub usage is aggregated over a more diverse set of OD pairs, reducing the risk of overfitting to noise either in pre-estimated utilities or in observed hub counts.

These experimental results suggest that when the proportion of hub trips in the system remains low and the number of OD pairs is large (Scenario 3), an estimation error within approximately ±10% in observed hub usage is unlikely to significantly impact the calibrated behavioral parameters. Nevertheless, the accuracy of the observed hub trip count remains an important factor influencing model outcomes.

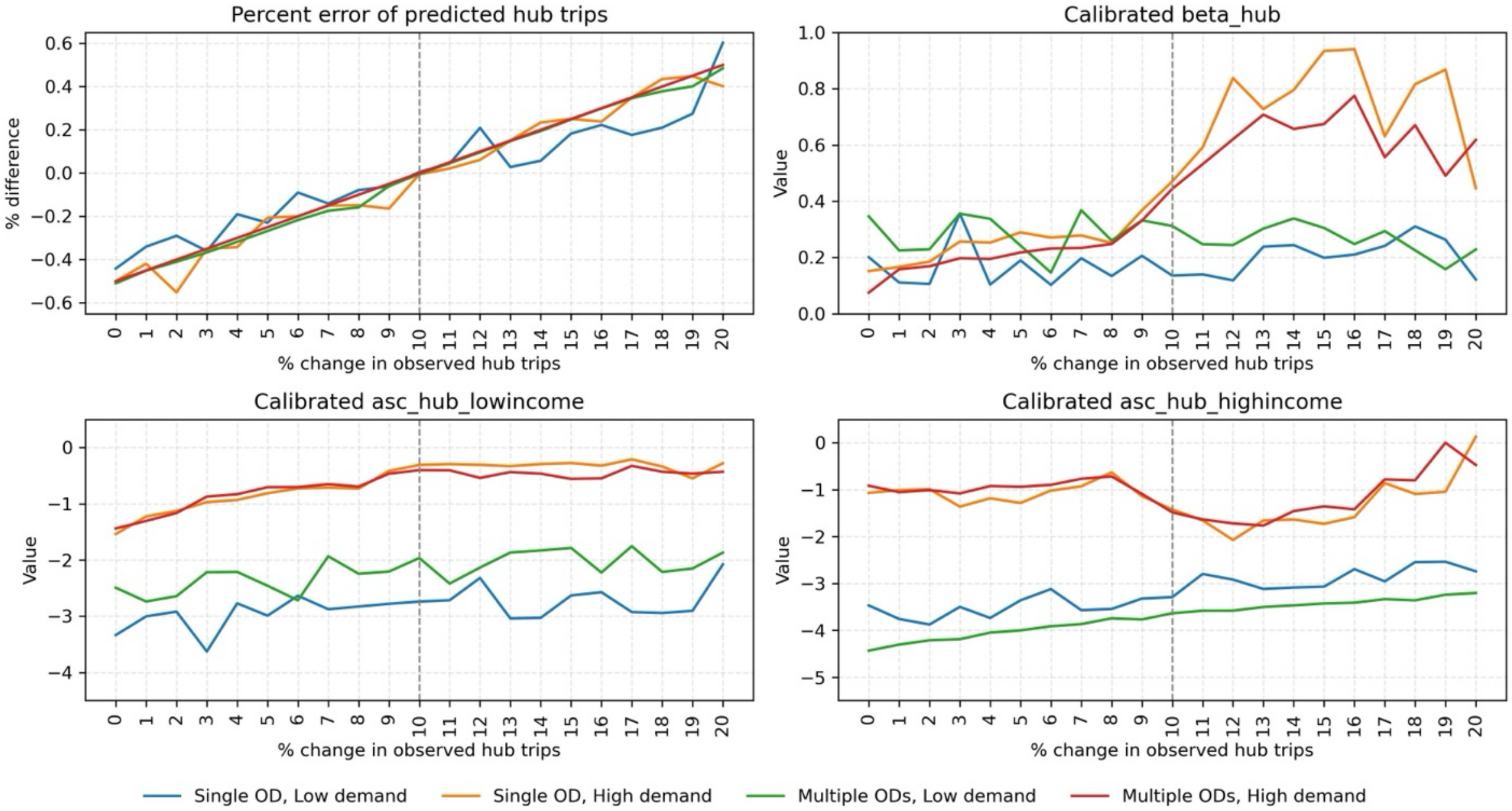

**Fig. 3.** Prediction error in trip counts and calibrated hub-related parameters corresponding to perturbations in observed hub trip.

### 3.3. Metrics for impact assessment and location selection

Following previous studies introduced in Section 2.1, we focus on several metrics for impact assessment and location selection. These metrics can be directly obtained from the mode choice model with calibrated hub-related parameters.

***Potential demand***: Estimated by identifying trips with OD pairs that fall within the threshold defined in Section 3.1.1. These are trips for which a hub-enabled multimodal option is feasible and competitive. The total potential demand reflects the number of such trips, as defined in Eq. (11).

$$PD_h = \sum_{t \in T} \sum_{i \in I} d_{h,i,t}\,, \quad \forall h \in H \tag{11}$$

where $PD_h$ denotes the potential trip demand for hub $h$, and $d_{h,i,t}$ denotes the number of trips made by population segment $i$ along OD pair $t$ that are potentially impacted by hub $h$.

***Mode share after hub demonstration***: This reflects the proportion of each traditional (unimodal) mode once multimodal trip options enter the choice set. Because a multimodal trip is physically composed of two unimodal legs (a first leg using mode $j_1$ from origin $O_t$ to hub $h_t$ and a second leg using mode $j_2$ from hub $h_t$ to destination $D_t$), its choice probability must be attributed back to the constituent modes before metrics like transit ridership or car VMT can be computed. We adopt a distance-based attribution, assigning the probability of each multimodal alternative to its two legs in proportion to the leg travel distances, as defined in Eq. (12).

$$P_{i,t}\big(j_1 \big| (j_1, j_2)\big) = \frac{l_{j_1, O_t \to h_t}}{l_{j_1, O_t \to h_t} + l_{j_2, h_t \to D_t}} P_{i,t}\left((j_1, j_2) \middle| J \cup J_{h_t}^{hub}\right), \quad \forall i \in I, (j_1, j_2) \in J_{h_t}^{hub}, t \in T \tag{12}$$

where $P_{i,t}(j_1|(j_1, j_2))$ denotes the share of multimodal alternative $(j_1, j_2)$ attributed to its first leg's mode $j_1$; $l_{j_1, O_t \to h_t}$ and $l_{j_2, h_t \to D_t}$ are the travel distances of the first and second legs, respectively; and $P_{i,t}\big((j_1, j_2)\big|J \cup J_{h_t}^{hub}\big)$ is the probability of choosing the multimodal alternative $(j_1, j_2)$ given the combined choice set of unimodal and multimodal alternatives. The share attributed to the second leg's mode $j_2$ is obtained symmetrically so that the two leg shares sum to the full probability of the multimodal alternative.

The total share of a traditional mode $j$ after hubs are introduced is then the sum of its purely unimodal share and the shares it receives from every multimodal alternative in which it serves as a leg, as defined in Eq. (13):

$$P_{i,t}\big(j \big| J \cup J_{h_t}^{hub}\big) = P_{i,t}(j|J^+) + \sum_{\substack{(j_1, j_2) \in J_{h_t}^{hub} \\ j \in \{j_1, j_2\}}} P_{i,t}\big(j_1 \big| (j_1, j_2)\big), \quad \forall i \in I, j \in J, t \in T \tag{13}$$

where $P_{i,t}\big(j\big|J \cup J_{h_t}^{hub}\big)$ is the share of mode $j$ for population segment $i$ along OD pair $t$ over the combined choice set; $P_{i,t}(j|J^+)$ is the probability of choosing traditional mode from the upper branch, and the summation term aggregates the attributed shares from Eq. (12) over all multimodal alternatives for which $j$ is one of the modes in the two legs from the lower branch.

***Increased transit ridership***: Quantified by comparing the predicted distribution of travel modes before and after introducing the mobility hub. For trips using mobility hubs, modal shares are proportionally weighted by segment distances to reflect partial contributions of each mode. The increased transit ridership is defined in Eq. (14).

$$TR_h = \sum_{t \in T} \sum_{i \in I} d_{h,i,t} \left(P_{i,t}\big(transit \big| J \cup J_{h_t}^{hub}\big) - P_{i,t}(transit|J)\right), \quad \forall h \in H \tag{14}$$

where $P_{i,t}\big(transit\big|J \cup J_{h_t}^{hub}\big)$ denotes the transit mode share after hubs are available, and $P_{i,t}(transit|J)$ denotes the transit mode share without hub demonstration.

***Reduced car VMT***: Measured by comparing the total vehicle-miles-traveled (VMT) by private cars before and after the introduction of the mobility hub. For each motorized, non-transit trip, network travel distances are estimated using the open-source map API. The reduction in car VMT reflects the environmental benefits brought by the hub, as shown in Eq. (15).

$$RV_h = \sum_{t \in T} \sum_{i \in I} l_t d_{h,i,t} \left( P_{i,t}\left(car \middle| J \cup J_{h_t}^{hub}\right) - P_{i,t}(car|J) \right), \quad \forall h \in H \tag{15}$$

where $l_t$ denotes driving network distance of OD pair $t$. $P_{i,t}\left(car \middle| J \cup J_{h_t}^{hub}\right)$ and $P_{i,t}(car|J)$ are probabilities of selecting private auto mode (including driving and carpool) before and after the hub demonstration. We further estimate carbon emissions by converting VMT using a factor from the U.S. Environmental Protection Agency (EPA), which average approximately 400 grams of $CO_2$ per mile for private vehicles (EPA, 2024).

***Increased consumer surplus***: Calculated as the difference in expected utility per trip before and after the demonstration of mobility hubs, converted into monetary terms using the estimated cost parameter from the mode choice model. This metric reflects the added value or satisfaction that travelers gain from having transfer options enabled by the mobility hub, which is defined in Eqs. (16) – (17).

$$CV_h = \sum_{t \in T} \sum_{i \in I} d_{h,i,t} \left( CS_{i,t}(J^+) - CS_{i,t}(J) \right), \qquad \forall h \in H \tag{16}$$

$$CS_{i,t}(K) = \frac{1}{\alpha_{i,t}} \left( \ln \sum_{j \in K} e^{V_{i,j,t}} + C \right), \qquad \forall i \in I, t \in T, K \in \{J^+, J\} \tag{17}$$

where $CS_{i,t}(J^+)$ and $CS_{i,t}(J)$ are the consumer surplus (CS) (McConnell, 1995) of population $i$ along OD pair $t$ given choice sets $J^+$ and $J$. $C$ is an arbitrary constant specific to $i$ and $t$. $\alpha_{i,t}$ is a parameter used to covert CS into monetary units (dollars), and is set equal to the travel cost parameter. $CV_h$ denotes the increased CS—also called compensating variation (Small & Rosen, 1981)—brought by the mobility hub.

## 4. Case study in the Capital District, NY

In this section, we apply the proposed framework to a case study in Capital District, NY. The Capital District Transportation Authority (CDTA) launched the Capital Region Mobility Hubs Demonstration Project, which ran from April 2022 to June 2024, to pilot two mobility hubs—one close to the UAlbany Downtown Campus and the other in Downtown Cohoes. Both hubs have been operational for several months, providing real-world usage data and the opportunity to conduct on-site surveys. Fig. 4 shows the locations of the two mobility hubs and trip OD pairs potentially impacted.

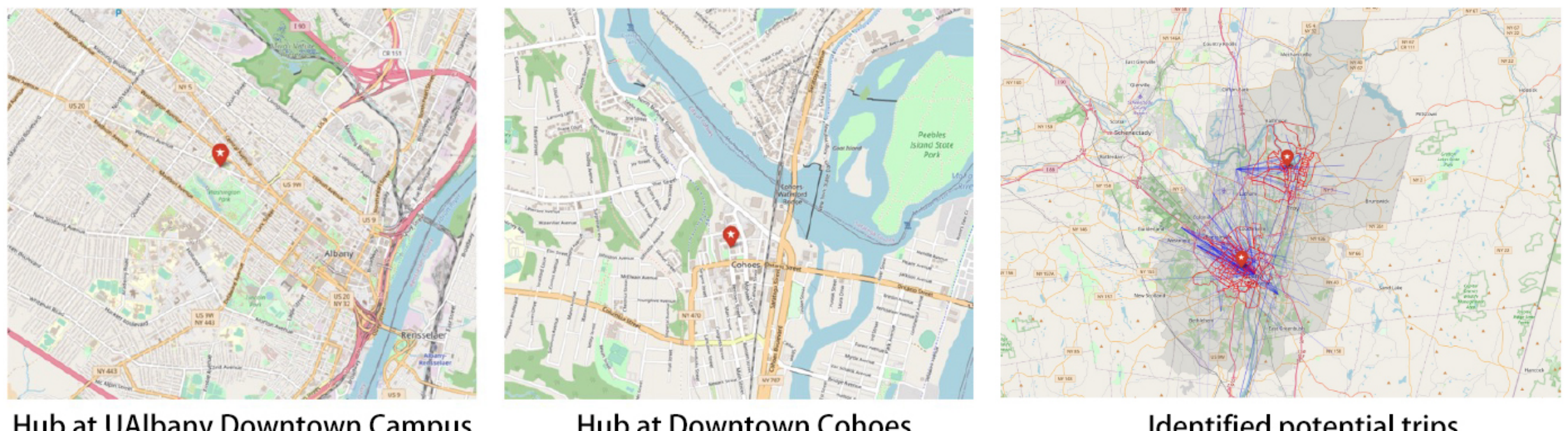

**Fig. 4.** Location of the two mobility hubs and trips potentially impacted.

### 4.1. Data collection

#### *4.1.1. Onsite hub survey*

We designed an on-site survey to collect mobility hub usage data, which was conducted by CDTA and yielded 40 complete responses between October and December 2023. The first section of the survey includes eight questions about a real trip involving the mobility hub, focusing on respondents' transportation choices and their reasons for using specific modes available at the hub. The second section consists of seven questions about personal information, used to identify population segments with unique sensitivity to travel time, cost, and modes. All questions used multiple-choice formats with an optional "prefer not to say" response. A full version of the questionnaire is available at: https://nyu.qualtrics.com/jfe/form/SV_9sHTmTbDu18ORng.

Fig. 5 summarizes key statistics from the 40 survey responses collected at the two pilot sites. Of these, 22 were collected at the UAlbany Downtown Campus and 18 in Downtown Cohoes. No respondent reported using both hubs within the same trip. As shown in panel (a), the most common trip purposes were commuting to or from school (30%) and shopping or errands (27.5%), followed by commuting to or from work (22.5%), other personal purposes (17.5%), and leisure or exercise (7.5%). Regarding trip frequency (panel b), a majority of respondents used the mobility hub regularly, with 35% reporting 1–3 days per week and 30% 4 or more days per week. Panels (c) and (d) show that the bus was by far the dominant travel mode both to and from the hub (over 60% of all trips), followed by walking and car share. Panel (e) reveals that 62.5% of respondents had made similar trips before the hub demonstration. Among these respondents, over 40% relied on private cars prior to the hub implementation, indicating a potential mode shift from car-based travel to shared and active modes through the hub deployment.

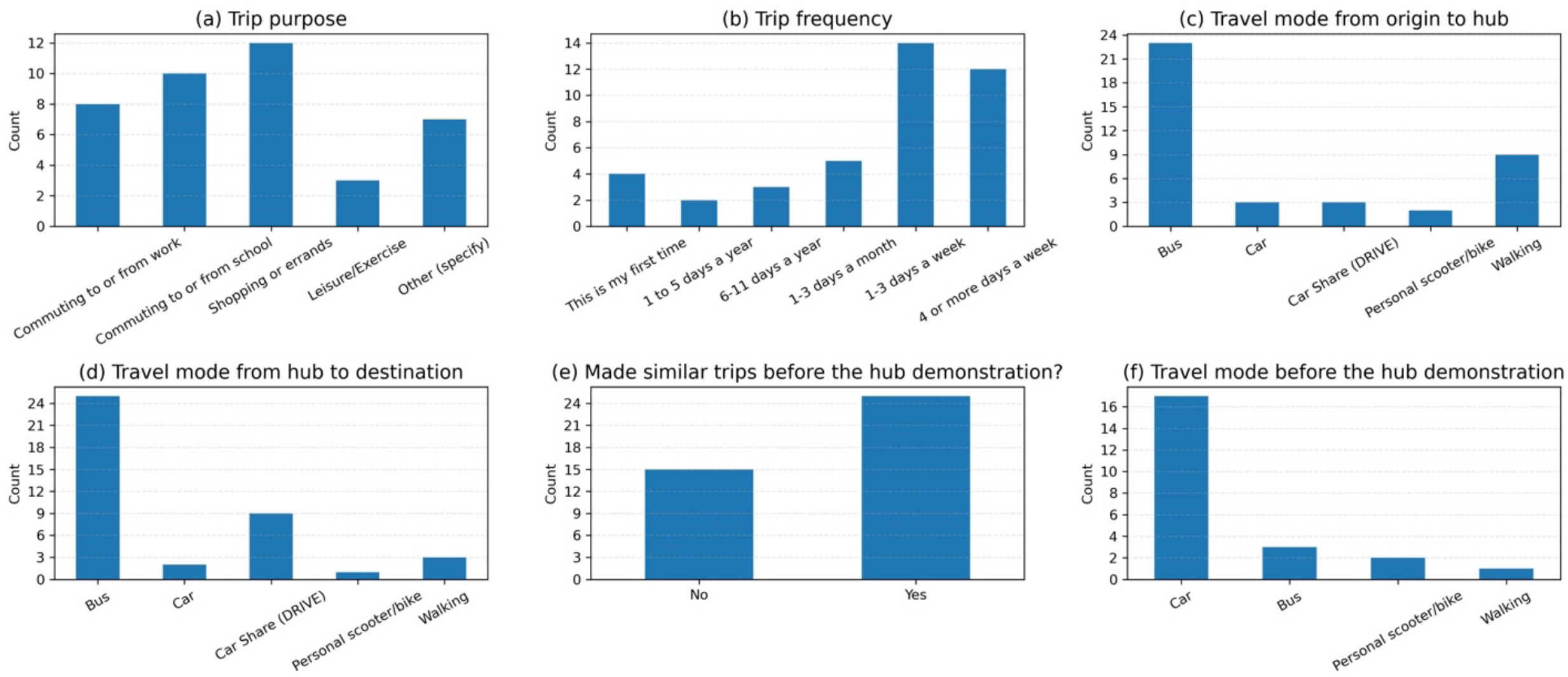

**Fig. 5.** A summary of survey results. In panels (c), (d), and (e), DRIVE is a car share service provided by the CDTA at a fare rate of $5 per hour.

*4.1.2. Synthetic trips and pre-estimated mode choice model*

We used a New York State mode choice model developed using Replica's synthetic trip data (Ren et al., 2025), publicly available at https://zenodo.org/records/19581095. The dataset contains over 50 million synthetic trips made by NYS residents on a typical weekday in 2019, which was generated through a combination of census data, mobile phone data, economic activity data, and built environment data (Replica Inc., 2024). Six trip modes are included: driving, public transit, on-demand auto, biking, walking, and carpool. Since the pre-estimated choice model is based on 2019 data, we compared the 2019 and 2023 Replica datasets to assess whether the underlying travel patterns remained stable. As shown in Table 3, transit and walking mode shares at both hubs increased slightly from 2019 to 2023, reflecting ongoing shifts toward more sustainable and active travel modes in the region. However, these changes are small in magnitude, and the relative ranking and dominance of driving mode are unchanged. To this end, the use of the 2019-based mode choice model remains reasonable for representing traveler preferences in our case.

**Table 3.** Number of potential trips and mode shares.

| | 2019 synthetic data | | 2023 synthetic data | |
|---|---|---|---|---|
| | UAlbany Hub | Cohoes Hub | UAlbany Hub | Cohoes Hub |
| Num. potential trips | 20,511 | 5,470 | 20,885 | 5,277 |
| Driving mode share | 67.28% | 56.14% | 67.62% | 60.05% |
| Transit mode share | 3.09% | 8.55% | 4.13% | 9.15% |
| On-demand mode share | 2.35% | 0.59% | 3.26% | 0.55% |
| Biking mode share | 0.74% | 0.38% | 0.64% | 0.31% |
| Walking mode share | 8.70% | 16.62% | 6.70% | 14.60% |
| Carpool mode share | 17.85% | 17.73% | 17.64% | 15.33% |

Note: Potential trips were identified using the approach in Section 3.1.1 and visualized in Fig. 4. The threshold for identification is set to 1.18 based on our survey responses.

The pre-estimated model is a nonparametric extension of the Berry–Levinsohn–Pakes (BLP) model (Berry et al., 1995), which serves as a benchmark for market-level discrete choice modeling. It achieves an out-of-sample prediction accuracy of 81.78%, substantially outperforming the standard BLP model on the same dataset (Ren et al., 2025). For each trip OD pair and population segment, the estimated $\beta_{i,t}$ contains twelve taste parameters: travel time for auto; in-vehicle time for public transit; access time for public transit; egress time for public transit; number of transfers for public transit; travel time for non-vehicle modes; trip monetary cost; and mode-specific constants for driving, public transit, on-demand auto, biking, and walking. Carpool is set as the reference mode. Unlike the traditional BLP formulation, this model estimates a unique set of parameters for each census block group OD pair, further segmented into four mutually exclusive population groups: non-low-income, low-income, seniors, and students. These segments are defined as follows: students include individuals currently enrolled in schools, colleges, or universities; seniors are those aged 65 or older; and the remaining population is classified into low-income and non-low-income groups based on the U.S. Federal Poverty Guidelines[1].

## 4.2 Mode choice model

### *4.2.1. Observed hub trip counts*

Obtaining the observed number of trips using each mobility hub ($Num_h$) presents a key challenge due to limited data availability. While daily trip counts were not feasible to collect during the hub survey, we leveraged a combination of on-site survey data and DRIVE backend data to approximate hub usage levels.

We use the Car Share (DRIVE) mode as a bridge between two data sources: the on-site survey, which records respondents' mode choices at each hub, and the DRIVE operator's backend system, which provides a ground-truth count of DRIVE trips originating from the Cohoes hub (DRIVE service is unavailable at the UAlbany hub). Because only DRIVE counts are directly observed, we anchor the estimation to DRIVE and scale all other quantities from it.

Let $N = 40$ denote the total number of survey responses over a window of $D_s = 61$ days, with $n_{UAlbany} = 22$ and $n_{Cohoes} = 18$ responses associated with the two hubs, respectively. Let $M_{Cohoes} = 28$ denote the number of DRIVE trips over a window of $D_O = 30$ days in October. From the survey, we estimate the overall DRIVE usage proportion $\hat{p}_{DRIVE}$. The daily trip volume at the Cohoes hub is then recovered as Eq. (18).

$$\widehat{Num}_{Cohoes} = \frac{M_{Cohoes}}{\hat{p}_{DRIVE} \cdot D_O} \tag{18}$$

from which an effective survey sampling rate $\hat{r}_s = n_{Cohoes}/(\widehat{Num}_{Cohoes} \cdot D_s)$ is inferred. Assuming the same sampling rate applies at the UAlbany hub, we invert the relationship to recover $\widehat{Num}_{UAlbany}$, as shown in Eq. (19).

$$\widehat{Num}_{UAlbany} = \frac{n_{UAlbany}}{\hat{r}_s \cdot D_s} = \widehat{Num}_{Cohoes} \cdot \frac{n_{UAlbany}}{n_{Cohoes}} \tag{19}$$

Because the survey yielded a limited number of responses and the estimation of $\widehat{Num}_{Cohoes}$ and $\widehat{Num}_{UAlbany}$ rely on simplified proportional relationships between backend trip counts, survey-reported shares, and total trip potentials rather than direct observations, a single point estimate risks conveying false precision. To quantify this uncertainty, we employ a bootstrap approach over the 40 survey respondents: in each of $B = 1{,}000$ replications, we draw 40 responses with replacement and recompute the full pipeline, yielding 1,000 bootstrap replicates. Fig. 6 shows the resulting distributions of daily hub trip counts. The UAlbany hub yields a mean of 9.78 trips/day, with a 95% confidence interval (CI) of [3.20, 27.99]. The Cohoes hub results in a mean of 7.10 trips/day with a 95% CI of [3.93, 14.93]. The confidence intervals span a wide range, roughly 0.5×–2× the mean for Cohoes hub and 1/3×–3× the mean for UAlbany. However, both distributions are right-skewed with a long upper tail, meaning the bulk of the probability mass remains concentrated near the mean and the central estimates are reasonably representative despite the apparent spread.

[1] https://aspe.hhs.gov/topics/poverty-economic-mobility/poverty-guidelines

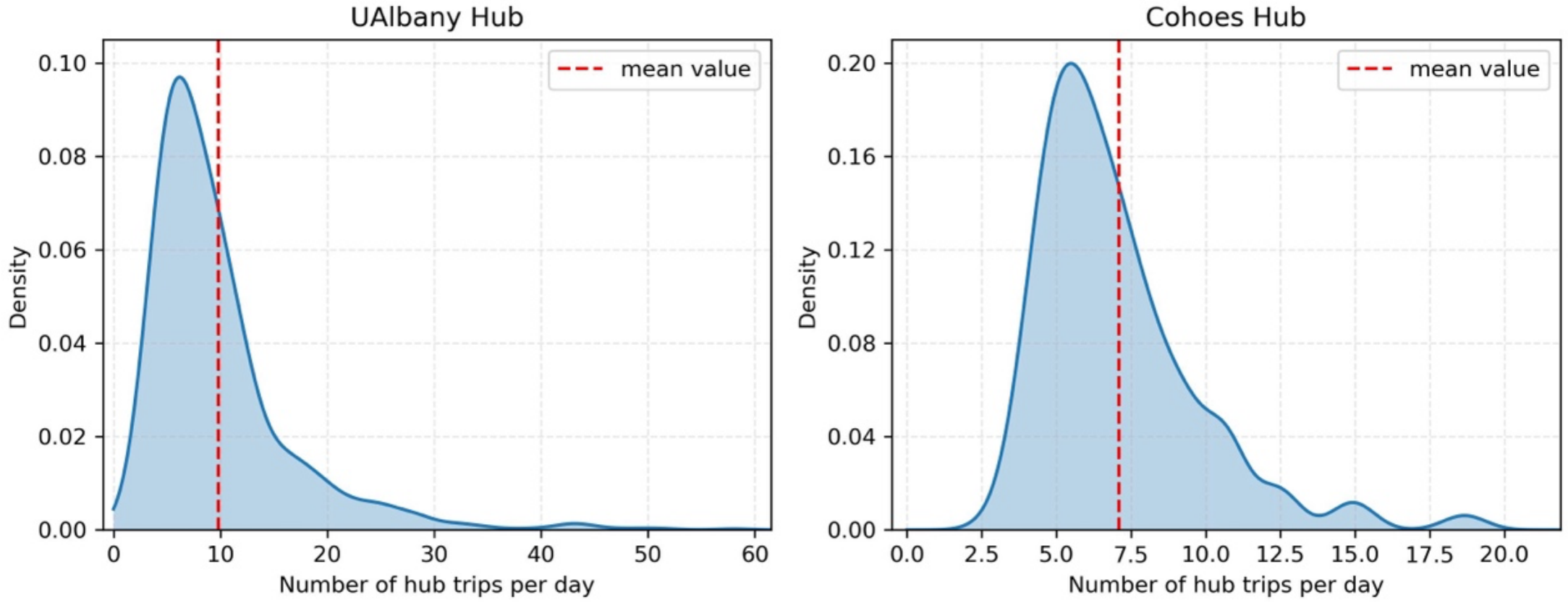

**Fig. 6.** Bootstrap distributions of daily hub trip counts.

Every downstream quantity, including calibrated hub-utility coefficients (Section 4.2.2), predicted mode shifts, VMT impacts, $CO_2$ reductions, and consumer-surplus gains (Section 4.3), is computed per bootstrap replicate rather than solely from the point estimate. All reported results therefore carry a bootstrap mean and 95% confidence interval, ensuring that survey-sample uncertainty is honestly reflected in the final estimates.

*4.2.2. Hub-related parameter calibration*

From the survey, we identified 16 available mode combinations at the Cohoes hub and nine combinations at the UAlbany hub (since DRIVE is not available there), as listed in Table 4. The travel time matrices for driving, biking, and walking were obtained from OpenStreetMap via the OSRM [2] package. Transit travel times were derived from CDTA's GTFS data [3] and OpenTripPlanner[4]. Fare tables for bike share[5] and car share[6] services were used to estimate the travel costs associated with shared mobility options.

**Table 4.** Available mode combinations at the Cohoes hub and UAlbany hub.

| Mode combinations at the Cohoes hub | Mode combinations at the UAlbany hub |
| --- | --- |

[2] https://github.com/Project-OSRM/osrm-backend
[3] https://www.cdta.org/gtfs
[4] https://www.opentripplanner.org/
[5] https://www.cdta.org/news/season-8-cdphp-cycle
[6] https://drivecdta.org/

| | |
|---|---|
| Bus—Bus, Bus—Car Share, Bus—Walking,<br>Bus—Bike Share, Car Share—Bus,<br>Car Share—Walking, Car Share—Bike Share,<br>Car—Bus, Car—Walking, Car—Car Share,<br>Car—Bike Share, Bike Share—Bus,<br>Bike Share—Walking, Bike Share—Car Share,<br>Walking—Bus, Walking—Car Share | Bus—Bus , Bus—Walking, Bus—Bike Share,<br>Car—Bus, Car—Walking, Car—Bike Share,<br>Bike Share—Bus, Bike Share—Walking,<br>Walking—Bus |

Given the sparse data, we adopt the simplified calibration approach proposed in Section 3.1.3. The calibration results are shown in Table 5. The predicted hub trip counts closely match the observed values, demonstrating that the proposed calibration procedure performs well in aligning model outputs with empirical usage data.

**Table 5.** Calibrated hub-specific parameters and prediction results (bootstrap mean with standard deviation and 95% CI from 1,000 replications).

| | Mean value | Standard deviation | 95% CI |
|---|---|---|---|
| **Calibrated parameters** | | | |
| $\beta_{hub}$ | 0.1278 | 0.0316 | [0.0179, 0.2189] |
| $asc_{notlowincome}^{hub}$ | –4.9598 | 0.5659 | [–6.3419, –3.6250] |
| $asc_{lowincome}^{hub}$ | –7.5466 | 1.3014 | [–9.2821, –5.2749] |
| $asc_{hub,senior}$ | –2.9584 | 0.1401 | [–3.3097, –2.4322] |
| $asc_{hub,student}$ | –4.4411 | 0.2941 | [–5.3335, –3.5228] |
| **Predicted hub trips (trips/day)** | | | |
| UAlbany hub (observed) | 9.78 | 6.79 | [3.20, 27.99] |
| UAlbany hub (predicted) | 9.89 | 6.84 | [3.20, 29.04] |
| Cohoes hub (observed) | 7.10 | 2.86 | [3.93, 14.93] |
| Cohoes hub (predicted) | 6.98 | 2.82 | [3.86, 12.67] |

### 4.3. Impact assessment of implemented hubs

Building on the calibrated model in the preceding sections, we now turn to the question of what the deployed mobility hubs actually deliver to the surrounding travel system. The impact assessment is organized around three complementary dimensions that, together, characterize hub performance from a traveler-centric perspective. Reporting each effect with its associated 95% confidence interval allows us to distinguish robust behavioral responses from differences that fall within the noise of stochastic demand realizations.

#### *4.3.1. Impacts on mode shift*

Table 6 shows the predicted mode shifts for potential trips after the demonstration of the mobility hubs in UAlbany and Cohoes. In general, our model predicts an average of 9.89 multimodal trips/day at the UAlbany hub (95% CI: [3.20, 29.04]). Most of these trips were shifted from traditional driving trips (–6.65 trips/day, 95% CI: [–19.54, –2.15]), with additional contributions from carpool (−1.76 [−5.18, −0.57]) and walking (−0.86 [−2.52, −0.28]). The average increase in bus ridership (transit) is 4.62 − 0.31 = 4.31 trips/day, combining the new multimodal transit trips (4.62 [1.60, 13.31]) with the small reduction in unimodal transit (−0.31 [−0.90, −0.10]).

As for the Cohoes hub, our model predicts an average of 6.98 trips/day using the hub (95% CI: [3.86, 12.67]). About half of these trips were shifted from driving trips (−3.92 trips/day, 95% CI: [−7.11, −2.17]), followed by carpool (−1.24 [−2.25, −0.68]) and walking (−1.16 [−2.10, −0.64]). The average increase in bus ridership was 3.46 − 0.60 = 2.86 trips/day, combining the multimodal transit gain (3.46 [1.87, 7.23]) with the unimodal transit loss (−0.60 [−1.08, −0.33]).

**Table 6.** Number of daily trips by travel mode with and without the hub deployment. Values are reported as mean [95% confidence interval].

| | Without Hub (Unimodal) | With Hub (Unimodal) | Num. Change (Unimodal) | With Hub (Multimodal) |
|---|---|---|---|---|
| **UAlbany hub** | | | | |
| Driving | 13,798 | 13,791.60 [13,779.10, 13,795.88] | −6.65 [−19.54, −2.15] | 2.04 [0.68, 5.64] |
| Transit | 635 | 634.30 [633.57, 634.56] | −0.31 [−0.90, −0.10] | 4.62 [1.60, 13.31] |
| On-demand auto | 482 | 481.43 [481.15, 481.50] | −0.23 [−0.68, −0.08] | — |
| Biking | 151 | 150.89 [150.69, 150.96] | −0.07 [−0.21, −0.02] | 1.66 [0.36, 5.54] |
| Walking | 1,785 | 1,783.75 [1,781.15, 1,784.40] | −0.86 [−2.52, −0.28] | 1.57 [0.27, 6.47] |
| Carpool | 3,660 | 3,658.14 [3,653.74, 3,659.68] | −1.76 [−5.18, −0.57] | — |
| Total | 20,511 | 20,501.11 [20,481.96, 20,507.80] | –9.89 [–29.04, –3.20] | 9.89 [3.20, 29.04] |
| **Cohoes hub** | | | | |
| Driving | 3,070 | 3,066.14 [3,062.51, 3,067.67] | −3.92 [−7.11, −2.17] | 0.84 [0.49, 1.39] |
| Transit | 467 | 466.74 [466.34, 466.88] | −0.60 [−1.08, −0.33] | 3.46 [1.87, 7.23] |
| On-demand auto | 33 | 32.46 [32.44, 32.47] | −0.04 [−0.08, −0.02] | — |
| Biking | 21 | 20.62 [20.58, 20.63] | −0.03 [−0.05, −0.01] | 0.50 [0.28, 0.85] |
| Walking | 909 | 907.76 [906.18, 908.85] | −1.16 [−2.10, −0.64] | 0.97 [0.48, 1.64] |
| Carpool | 970 | 968.29 [967.12, 968.93] | −1.24 [−2.25, −0.68] | 1.20 [0.64, 2.33] |
| Total | 5,470 | 5,463.02 [5,457.33, 5,466.14] | –6.98 [–12.67, –3.86] | 6.98 [3.86, 12.67] |

*4.3.2. Impacts on VMT and carbon emission*

Table 7 shows the predicted impacts on VMT and carbon emission. In general, the two hubs were projected to reduce an average VMT by $43.29 + 32.12 = 75.41$ miles per day, or 27.52 thousand miles in a year. The corresponding greenhouse gas (GHG) emissions reduction was $17.31 + 12.86 = 30.17$ kg per day, or $6.32 + 4.69 = 10.91$ metric tons of carbon emissions (MTCE) per year, which is close to the total carbon emissions of energy use of two households in a year[7]. In other words, there is potential for each mobility hub installment to reduce carbon emissions equivalent to one household's energy use each, although effectiveness varies by case.

**Table 7.** Impacts on VMT and carbon emission. Values are reported as mean [95% confidence interval].

| | VMT without hub (miles/day) | VMT with hub (miles/day) | VMT changes (miles/day) | Reduced $CO_2$ (kg/day) | Reduced $CO_2$ (tons/year) |
|---|---|---|---|---|---|
| **UAlbany hub** | | | | | |
| Driving | 100,550 | 100,518 [100,446, 100,540] | –32.00 [–103.59, –9.05] | –12.80 [–41.43, –3.62] | –4.67 [–15.12, –1.32] |
| Carpool | 24,509 | 24,495 [24,470, 24,506] | –11.29 [–39.22, –2.64] | –4.51 [–15.69, –1.05] | –1.65 [–5.73, –0.38] |
| Total | 125,059 | 125,016 [124,916, 125,047] | –43.29 [–142.81, –11.69] | –17.31 [–57.12, –4.67] | –6.32 [–20.85, –1.71] |
| **Cohoes hub** | | | | | |
| Driving | 28,573 | 28,541 [28,414, 28,570] | –32.48 [–58.71, –2.48] | –13.00 [–23.49, –0.99] | –4.75 [–8.57, –0.36] |
| Carpool | 7,311 | 7,311 [7,309, 7,312] | 0.36 [–1.56, 1.44] | 0.14 [–0.62, 0.57] | 0.05 [–0.23, 0.21] |
| Total | 35,884 | 35,852 [35,824, 35,883] | –32.12 [–60.27, –1.04] | –12.86 [–24.11, –0.42] | –4.69 [–8.80, –0.15] |

*4.3.3. Impacts on consumer surplus*

Table 8 shows the increase of consumer surplus brought by the two mobility hubs. In general, the hub in UAlbany Downtown Campus brought an average consumer surplus gain of $0.1887/trip (95% CI [0.0985, 0.2572]), amounting to $3,870/day in aggregate (95% CI [2,020, 5,276]). The hub in Downtown Cohoes brought an average consumer surplus gain of $0.3273/trip (95% CI [0.2198, 0.3781]), amounting to $1,790/day in aggregate (95% CI [1,202, 2,068]). The interpretation is that having the mobility hub created economic value for travelers, equivalent to that dollar amount for each of their trips whether or not they used the hub.

The hub service fare for bus line is generally $1.50 per trip (without any discount considered). The added value per trip is around $0.20-$0.30, which is at about one fifth to one seventh of the full bus fare. Considering that user preferences are quite diverse, various pricing policies could be designed to balance the added value and charged fare per trip.

**Table 8.** Impacts on consumer surplus of potential trips. Values are reported as mean [95% confidence interval].

| | UAlbany hub | Cohoes hub |
|---|---|---|
| Increased consumer surplus ($/trip) | 0.1887 [0.0985, 0.2572] | 0.3273 [0.2198, 0.3781] |
| Number of potential trips per day | 20,511 | 5,470 |
| Total CV brought by the hub ($/day) | 3,870 [2,020, 5,276] | 1,790 [1,202, 2,068] |

## 4.4. Evaluation of potential hub locations

*4.4.1. Potential hub locations*

We further demonstrate how our calibrated model can support the identification of suitable locations for future mobility hub deployment. As mentioned in previous studies, public transport stops are initially considered as candidate hubs (Xanthopoulos et al., 2024). To this end, we treat existing CDTA bus stops in the Albany-Schenectady-Troy Metropolitan Statistical Area (MSA) as potential hub locations. To avoid redundancies, bus stops located within 200 meters of each other are clustered into a single hub candidate. This process results in a total of 1,100 unique hub candidates across the region.

[7] https://www.epa.gov/energy/greenhouse-gas-equivalencies-calculator-calculations-and-references

To reflect real-world service availability, we further differentiate these candidates based on proximity to CDTA park-and-ride (P+R) facilities[8]. For hub locations within 500 meters of a CDTA P+R lot, car-share service is assumed to be available, similar to the configuration observed at the Cohoes hub. For all other candidates, only bike share standard travel options are assumed, as in the case of the UAlbany hub. Fig. 7 displays the spatial distribution of these 1,100 hub candidates across the Albany-Schenectady-Troy MSA, demonstrating broad geographic coverage across both urban and suburban areas.

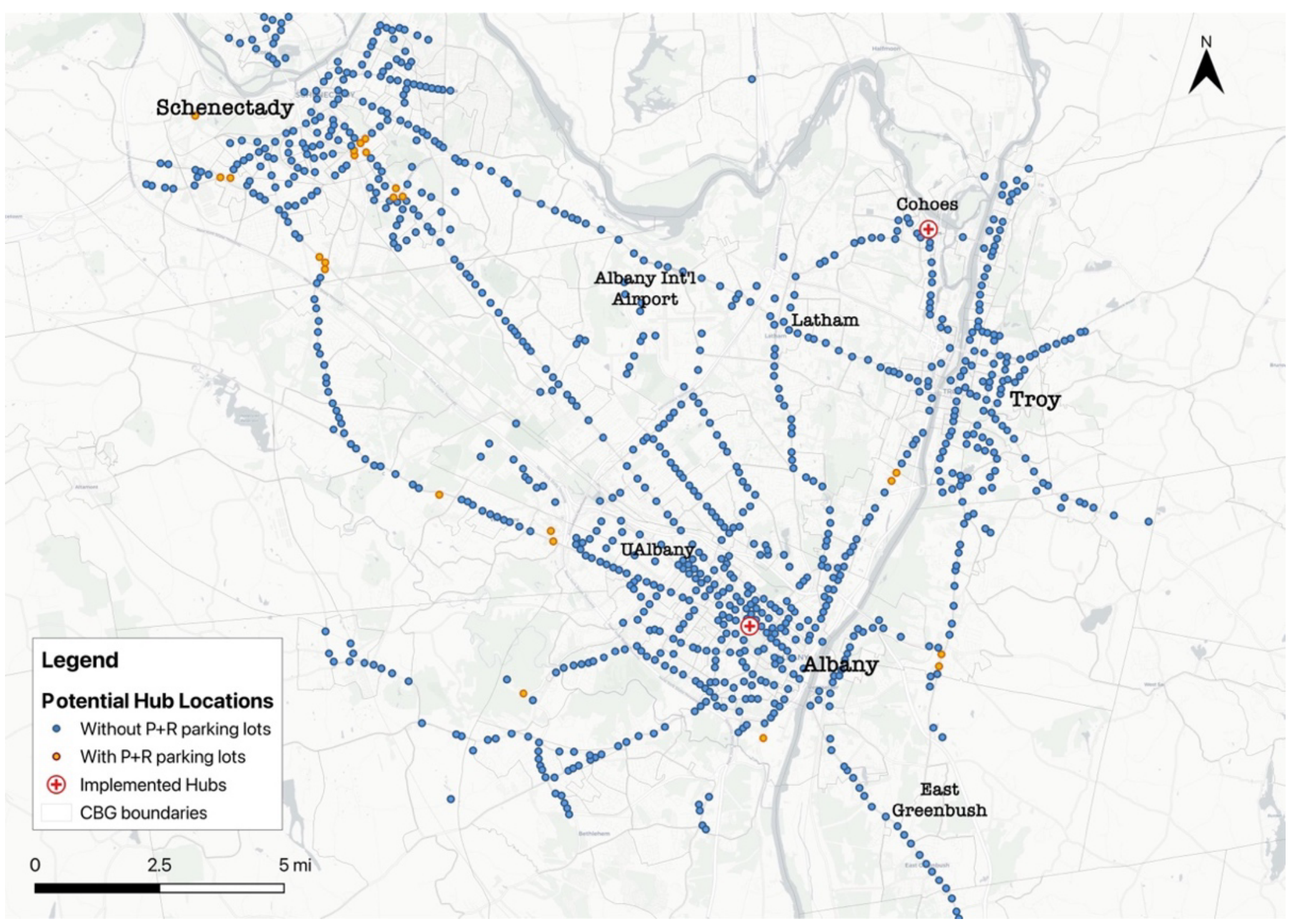

**Fig. 7.** Potential hub locations in the Albany-Schenectady-Troy MSA.

*4.4.2. Evaluation regarding key metrics*

Fig. 8 summarizes the descriptive statistics for all 1,100 potential mobility hub locations in the Albany-Schenectady-Troy MSA. Four performance metrics defined in Section 3.3 are reported: potential demand, increased transit ridership, reduced car VMT, and increased trip consumer surplus. All values shown are bootstrap means obtained from the pipeline described in Section 4.2. Potential hub demand varies widely, ranging from 308 to 27,921 potential trips per day, with a mean of 11,109 and standard deviation of 5,847. Transit ridership increase ranges from near zero to about 20 additional trips per day, with a mean of 5.22 and standard deviation of 4.82. Reductions in car VMT span from 0.10 to 410.77 miles per day, with a mean of 115.15 and standard deviation of 99.63. Total trip CS increases range from $10 to $26,371 per day, with a mean of $5,284 and a

[8] https://www.cdta.org/park-ride

standard deviation of $5,537. The histograms illustrate skewed distributions across all metrics, where a small number of hub locations deliver significantly higher benefits. Red vertical lines highlight the UAlbany and Cohoes hubs for comparison, showing both fall within the mid-to-upper percentiles across all four indicators.

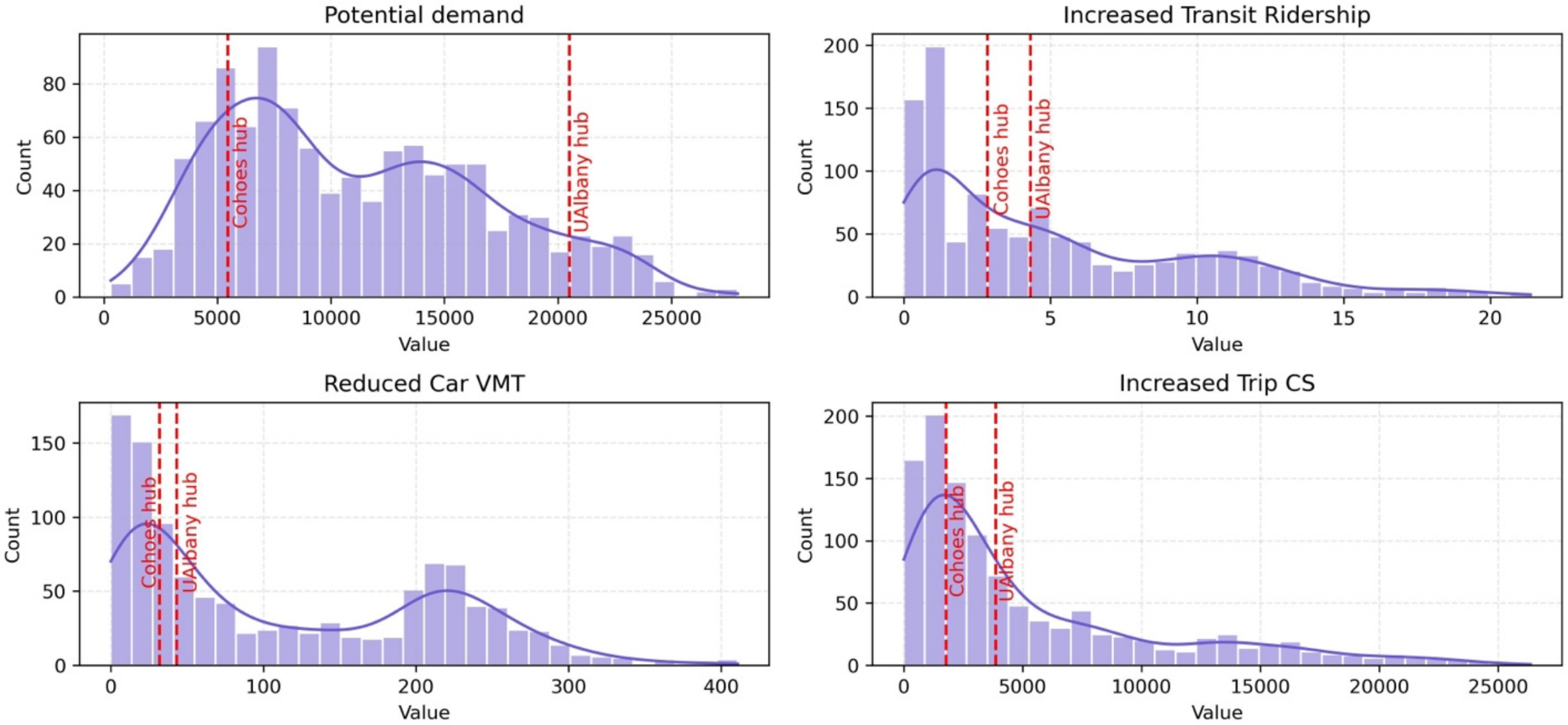

**Fig. 8.** Statistical distribution of candidate evaluation metrics.

Fig. 9 presents the spatial distribution of the four evaluation metrics for the 1,100 candidate locations. As shown in Panel (a), potential demand is concentrated in dense urban cores such as downtown Albany, Schenectady, and Troy, where overall travel activity is highest. However, Panels (b) and (c) reveal a different pattern: the most substantial increases in transit ridership and reductions in car VMT are found at hubs located around urban edges or along specific high-frequency bus corridors. This probably reflects a "P+R" usage pattern, where travelers switch from driving to transit at the urban fringe. Panel (d) highlights locations with the greatest increase in consumer surplus, which tends to occur along major intercity bus routes and near the UAlbany Downtown Campus, likely due to a combination of high travel demand and limited existing service. Overall, the results indicate that mobility hubs functioning as "P+R" connectors between cities have the greatest behavioral impacts. While the two implemented hubs (UAlbany and Cohoes) show significant benefits, they rank only in the 20–40% range across most metrics, suggesting they are moderately impactful but not among the top-performing 20% of candidate sites. Although practical hub location selections require further consideration of implementation costs and feasibility, our behavior-oriented evaluation provides a well-defined range of promising candidate locations for future mobility hub deployment.

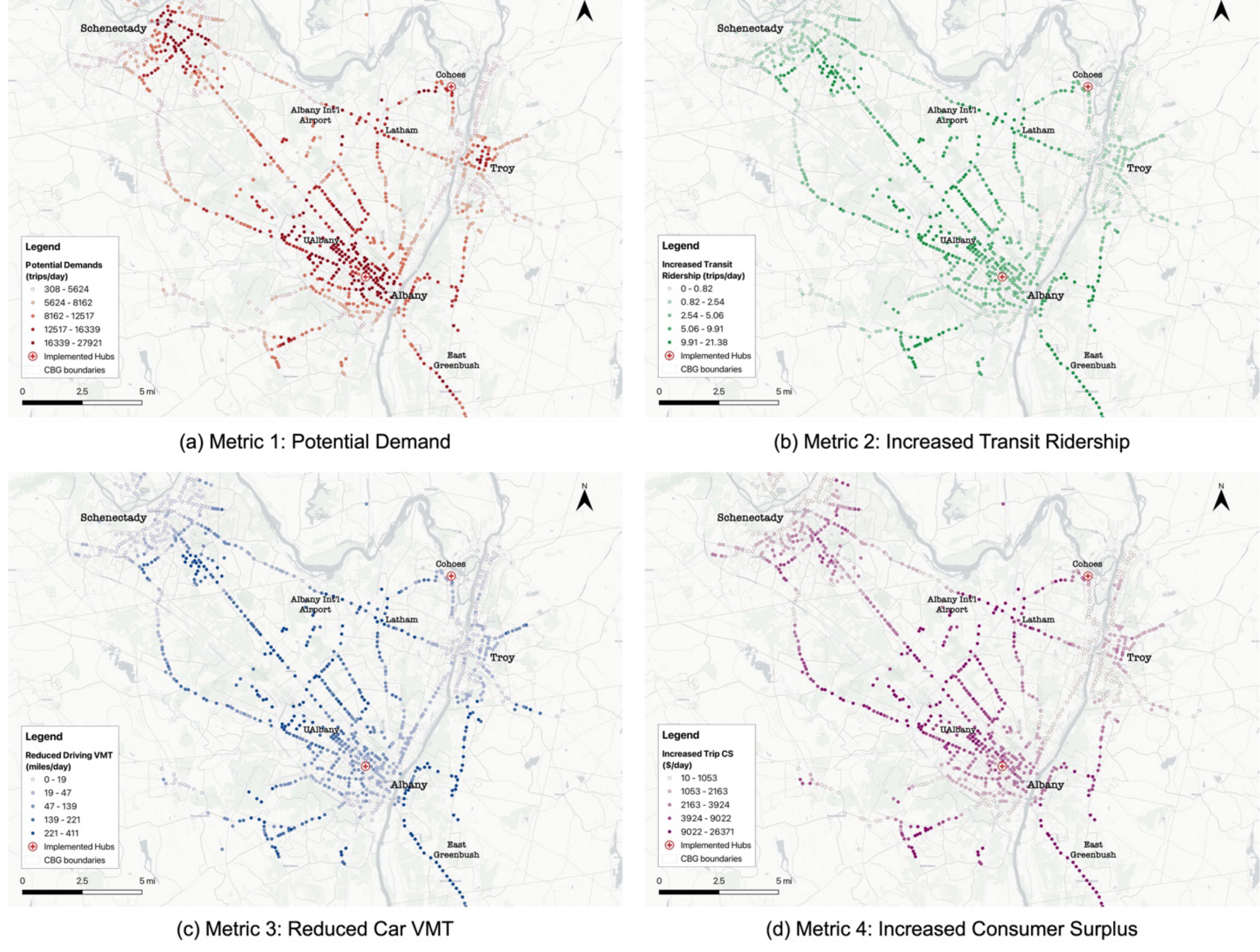

(a) Metric 1: Potential Demand

(b) Metric 2: Increased Transit Ridership

(c) Metric 3: Reduced Car VMT

(d) Metric 4: Increased Consumer Surplus

**Fig. 9.** Spatial distribution of candidate evaluation metrics.

## 5. Discussion and Conclusion

This study developed a data fusion approach that integrates observed mobility hub usage into a pre-estimated, large-scale mode choice model, enabling behaviorally grounded impact assessment without requiring full model re-estimation. Applied to a demonstration project in New York's Capital District, the framework quantified mode shift, VMT reduction, and consumer surplus gains at two implemented hubs and extended the analysis to 1,100 candidate locations across the Albany–Schenectady–Troy metropolitan area. In this section, we discuss generalizable insights that emerge from our findings and outline the key limitations of the current framework along with directions for future research.

### 5.1. Generalizable insights

Although our empirical analysis centers on two implemented hubs in New York's Capital District, the findings carry implications that extend beyond this specific setting. Our findings give rise to three insights that are relevant not only to mobility hub deployment but, more broadly, to the early-stage evaluation of emerging mobility innovations.

First, our framework provides a behaviorally consistent assessment that goes beyond simple before-and-after counts or independently derived performance metrics. Since mode shift, VMT and emissions reduction, transit ridership change, and consumer surplus all derive from a single nested logit structure embedded within a large-scale mode choice model, these metrics are internally linked and grounded in the same set of behavioral assumptions. The logit formulation also enables elasticity and welfare measures derived from random utility theory, that conventional aggregate calculations cannot produce. Equally important, hub-specific impacts are placed in context with statewide travel patterns: the 9.89 multimodal trips per day at the UAlbany hub and the 6.98 at Cohoes represent mode shifts from the 20,511 and 5,470 potential trips along the impacted OD pairs, rather than reported as isolated counts. This contextualization moves hub evaluation from isolated case reporting toward a comparable, system-level metric set.

Second, when applied across a wider set of candidate sites, the framework reveals that hub impact is heavily skewed across locations, and that site selection is an essential lever in deployment planning. Across the 1,100 candidates we evaluated in the Albany–Schenectady–Troy MSA, potential demand ranged from 308 to 27,921 trips per day, and consumer surplus gains from \$10 to nearly \$25,000 per day, with right-tailed distributions in which a small share of sites delivers the bulk of the benefit. Moreover, the highest-performing sites differ depending on the objective. Demand concentrates in dense urban cores; VMT reduction and transit ridership gains peak at urban peripheries and along high-frequency corridors that support park-and-ride patterns; and consumer surplus gains are greatest along intercity bus routes and in areas where high travel demand coincides with limited existing service. These objectives do not co-locate spatially, which means planners must be explicit about which outcome they are optimizing when selecting sites. The finding that the two implemented hubs rank only in the 20th–40th percentile across most metrics reinforces a more general point: pilot sites are typically selected on the basis of operational feasibility or institutional partnerships, and their observed performance should not be taken as indicative of what optimally sited hubs could deliver.

The third insight reaches beyond mobility hubs to the broader challenge of evaluating emerging mobility options before large-scale adoption data are available. Our data fusion approach demonstrates that synthetic-data-based choice models and small-sample on-site surveys are structurally complementary, not merely additive. Synthetic ICT-derived models provide regional scale, fine-grained OD coverage, and heterogeneous taste parameters, but cannot represent user preferences for modes that were absent when the data were generated. On-site surveys capture revealed behavioral responses to the new option, but lack the spatial and demographic reach needed for regional impact assessment. The data fusion strategy — freezing the pre-estimated taste parameters and identifying only the new-option-specific preference from marginal counts — exploits each data source for what it is uniquely suited to provide. This fusion logic is not specific to mobility hubs: it generalizes to the early-stage evaluation of any emerging service that introduces a new alternative into an existing transport system, such as autonomous shuttles, microtransit, and curb-pricing schemes, where deployment decisions have to be made before the longitudinal data become available.

### 5.2. Limitations and future work

The proposed framework rests on several simplifying assumptions that warrant discussion. First, due to the limited size of the on-site survey sample, we calibrate a reduced parameter set of $1 + |I|$ hub-specific quantities rather than the full specification of $1 + |I| \cdot |H| + |I| \cdot |J|^2$ parameters described in Section 3.1.3. This reduction implies two restrictions: (a) the transfer disutility within a hub is assumed to vary across population segments but remain constant across all mode-pair combinations, and (b) differences in unobserved factors across hubs are assumed to be small relative to those across population segments, such that all hubs share a common alternative-specific constant within each segment. While these restrictions are defensible at the current stage of deployment, they may become increasingly untenable as hub networks expand and richer usage data become available. Second, the extended mode choice model adopts a nested logit formulation that treats the hub alternative as parallel to and independent of the traditional modes in the upper branch. This structure imposes proportional substitution across the non-hub alternatives when a traveler switches to the hub option, which may not reflect actual behavioral patterns if, for instance, hub-enabled trips substitute disproportionately for driving rather than for walking or cycling. Although the survey responses reported in Section 4 (Fig. 5f) provide empirical support for approximately proportional substitution in our case study, alternative nesting structures or more flexible formulations should be explored as additional data become available.

These simplifications point toward concrete avenues for future methodological development. As hub networks mature and richer usage data accumulate, it will become feasible to calibrate the full parameter set, allowing transfer (dis)utilities to vary by mode-pair combination and alternative-specific constants to differ across hubs. Beyond richer parameterization, the nesting structure itself can be generalized. A particularly promising direction is the inverse product differentiation logit (IPDL) model (Fosgerau et al., 2024), which replaces the single nesting parameter with a set of $|D|$ parameters that capture correlations among alternatives along multiple dimensions, such as motorized versus non-motorized access or shared modes versus private modes. This flexibility is well suited to mobility hub contexts, where a hub-enabled trip may share attributes with driving along one dimension (e.g., door-to-door travel time) and with transit along another (e.g., shared-vehicle occupancy). Implementing the IPDL within the proposed data fusion framework would remain an open area for future research.

A further source of uncertainty lies in the estimation of observed hub trip counts $Num_h$, which serve as the primary calibration target in the data fusion procedure. Since direct counts of multimodal hub trips were not available, we inferred $Num_h$ through proportional relationships linking survey responses to backend car-share usage data (see Section 4.2.1). While this approach yields reasonable order-of-magnitude estimates, it introduces measurement error that propagates through the calibration and into all downstream impact metrics. The bootstrap procedure employed in our case study makes this uncertainty explicit: the 95% confidence interval for projected multimodal demand at the UAlbany hub spans from 3.20 to 29.04 trips per day around a mean of 9.89, a range exceeding 2.5 times the mean estimate; at the Cohoes hub, the interval of [3.86, 12.67] around a mean of 6.98 is narrower but still spans roughly 1.3 times the mean. These wide intervals reflect the compounded effect of small survey samples and indirect trip count inference. Narrowing them in future work will require richer ground-truth data sources, such as automated passenger counters at hub stations, passive mobility sensing, or fare-card transaction records. Moreover, hub usage is likely to evolve over time as travelers gain familiarity with the service and as broader hub deployment reshapes regional travel norms. A natural extension of this work would therefore be to periodically recalibrate the model parameters using updated trip counts, enabling agencies to track how traveler preferences toward hub-enabled modes are shaped by the cumulative expansion of the hub network over time.

**Acknowledgements**
The authors were partially supported by a NYSERDA-funded project, Capital Region Mobility Hubs (https://projects.cdta.org/capital-region-mobility-hubs), through a subcontract with Calstart. Data shared by Replica Inc. and CDTA are gratefully acknowledged. The views and opinions expressed in this article are solely those of the authors and do not necessarily reflect the official policy or position of CDTA, NYSERDA, Calstart, or Replica.

**CRediT authorship contribution statement**

The authors confirm contribution to the paper as follows. Xiyuan Ren: Conceptualization, Methodology, Analysis, Visualization, Draft writing, editing and review. Joseph Y. J. Chow: Conceptualization, Methodology, Data acquisition, Draft writing, editing and review.

**Declaration of generative AI and AI-assisted technologies in the writing process**

During the preparation of this work the authors used ChatGPT and Claude in order to improve readability and language. After using this tool/service, the authors reviewed and edited the content as needed and take full responsibility for the content of the published article.

**Declaration of competing interest**

The authors declared no potential conflicts of interest with respect to the research, authorship, and/or publication of this article.